\documentclass[%
 aip,
 amsmath,amssymb,
 reprint,%
groupedaddress
]{revtex4-1}

\usepackage{graphicx}
\usepackage{dcolumn}
\usepackage{bm}

\usepackage[utf8]{inputenc}
\usepackage[T1]{fontenc}
\usepackage{mathptmx}
\usepackage{etoolbox}
\usepackage{amsthm}
\usepackage{xcolor}
\usepackage{physics}
    
\makeatletter
\def\@email#1#2{%
 \endgroup
 \patchcmd{\titleblock@produce}
  {\frontmatter@RRAPformat}
  {\frontmatter@RRAPformat{\produce@RRAP{*#1\href{mailto:#2}{#2}}}\frontmatter@RRAPformat}
  {}{}
}%
\makeatother

\newtheorem*{theorem*}{Framework}

\begin{document}

\preprint{AIP/123-QED}
\title{A Dayem Loop Qubit Based on Interfering Superconducting Nanowires}
\email{cliffxs2@illinois.edu}
\author{Cliff Sun}
\email{bezryadi@illinois.edu}
\author{Alexey Bezryadin}%
\affiliation{Department of Physics, University of Illinois at Urbana-Champaign}

\begin{abstract}

We propose a qubit design based on two parallel superconducting nanowires (i.e., a "Dayem loop qubit"). The inclusion of two nanowires instead of one leads to the Little-Parks effect, which provides an option to control the qubit frequency in a wide range, which is possible because the vorticity remains fixed, since the rate of phase slips is exponentially low at low temperatures. Our key result is that even if each individual nanowire does not have a cubic nonlinearity in its current-phase relationship (CPR), quantum interference between two condensates, induced by a magnetic field, produces a cubic nonlinearity for the entire device. Our numerical simulations, supported by analytical solutions, show that such nonlinearity is sufficient to create functional transmon qubits based on pairs of thin homogeneous superconducting nanowires. We analyze both generic, i.e., cubic CPR, as well as more realistic microscopic CPR, having higher-order nonlinearities. For higher-order CPRs, we propose a simple power-law phenomenological approximation valid at very low temperatures, at which superconducting qubits normally operate.
\end{abstract}
\maketitle


\section{Introduction}

Quantum information processing has seen a rapid surge in interest due to its potential for dramatic speed-ups in optimization tasks \cite{shor_algorithm, grovers} and its ability to efficiently simulate high-dimensional quantum systems. Many major technology companies such as Microsoft, IBM, and Amazon are actively developing quantum computing resources, driving the need for scalable and robust quantum hardware.

Among the leading qubit architectures, superconducting qubits have emerged as the popular architecture for scalable quantum computing, but their reliance on millikelvin temperatures limits scalability \cite{nakamura-1999, collaborators-2024}. Devices capable of operating above 1~Kelvin could eliminate the need for dilution refrigerators and enable more practical large-scale systems. A key obstacle is the aluminum-based qubits used in most current quantum circuits. The small superconducting gap of aluminum restricts operation to below $\sim 200$~mK, where Bogoliubov quasiparticles cause significant decoherence \cite{10.21468/SciPostPhysLectNotes.31}.

Conventional metal--oxide--metal junctions (SIS junctions) also introduce parasitic capacitance, limiting the qubit operation frequency. High-temperature operation requires much higher frequencies to suppress thermal excitations. However, as the operating frequency is increased, the dielectric losses in the oxide barrier also increase, further reducing coherence times.

Several alternative strategies have been explored to overcome these constraints. One direction enhances conventional aluminum junctions by proximitizing them with higher-gap superconductors, such as niobium, thereby extending operation toward 1 Kelvin\cite{PhysRevApplied.21.024047}. A different route replaces aluminum-based elements entirely with nitride platforms, for example, NbN/AlN/NbN tunnel junctions \cite{wang-2026} or TiN nanowires \cite{purmessur-2025} acting as phase-slip junctions \cite{mooij-2006}, which inherently support higher-gap superconductivity and improved thermal robustness. Yet another approach introduced recently relies on superconductor-insulator quantum transition in NbN films \cite{bottcher2025transmonqubitrealizedexploiting}.

Here we develop an alternative idea, namely a fully metallic qubit that does not involve tunnel barriers, superconductor-semiconductor interfaces, quantum phase slips, or superconductor-insulator transitions. Our proposal is based on the superconducting quantum interference between two homogeneous fully metallic superconducting nanowires \cite{PhysRevB.82.134518, PhysRevB.94.165128, PhysRevLett.106.110502, vijay-2010, faramarzi-2021}, whose length $L$ is larger than its superconducting coherence length $\xi$. Unlike in tunnel junctions, the supercurrent in the nanowire does not flow through a lossy amorphous oxide; thereby, a major source of decoherence can be eliminated  \cite{PhysRevLett.93.077003}. Moreover, superconducting nanowires can be fabricated with diameters of only a few nanometers \cite{Bezryadin-2000}, a low critical current while using DNA as templates\cite{hopkins-2005}, and possess a high kinetic inductance \cite{PhysRevB.82.134518, PhysRevB.83.184503}, making them attractive for realizing sub-micron superconducting circuits. Also, the electric capacitance of a nanowire transmon qubit is much smaller than in traditional transmon based on SIS junctions since the distance between the superconducting banks (antennas) of the qubit equals the length of the wire, which is typically of the order of 100 nm or larger.

Models of single nanowire transmon qubits have previously been proposed \cite{PhysRevB.82.134518, faramarzi-2021, PhysRevLett.103.087003}. It was suggested that such qubits can operate in the W-band (75-110 GHz), thereby reducing the risk of decoherence and quasiparticle-induced thermodynamic fluctuations \cite{faramarzi-2021}. A DC current bias has been proposed to push single nanowire qubits to achieve anharmonicity similar to traditional transmon qubit \cite{PhysRevLett.103.087003}. But a major limitation of the single nanowire design is that its kinetic inductance becomes highly non-linear only at high supercurrents, whereas transmon qubits must exhibit a sufficiently high (at least one percent) nonlinearity when loaded with just one quantum of energy\cite{purkayastha2026tunableanharmonicitysninasnanowire}. The nanowires we consider here resemble Dayem bridges in the sense that they do not have any barriers and yet act as weak superconducting links because of the constriction of the condensate to a very narrow channel, practically of the order of 10 nm\cite{Bezryadin-2000,hopkins-2005,Bezryadin_book}. Nonlinear effects can also be increased by fabricating bilayer nanowires, which include normal metals driven superconducting by the proximity effect\cite{bilayer}.

Here we demonstrate that the nonlinearity can be substantially increased by connecting two nanowires in parallel and applying a magnetic field to induce a phase shift between them. Thus, we propose a qubit design composed of two parallel superconducting nanowires forming a superconducting quantum interference device (SQUID), i.e., a nanowire-based SQUID. We demonstrate that this architecture provides sufficient intrinsic nonlinearity to achieve the anharmonicity required for practical qubit operation.

 \begin{figure}[t]
     \centering
     \includegraphics[width=0.8\linewidth]{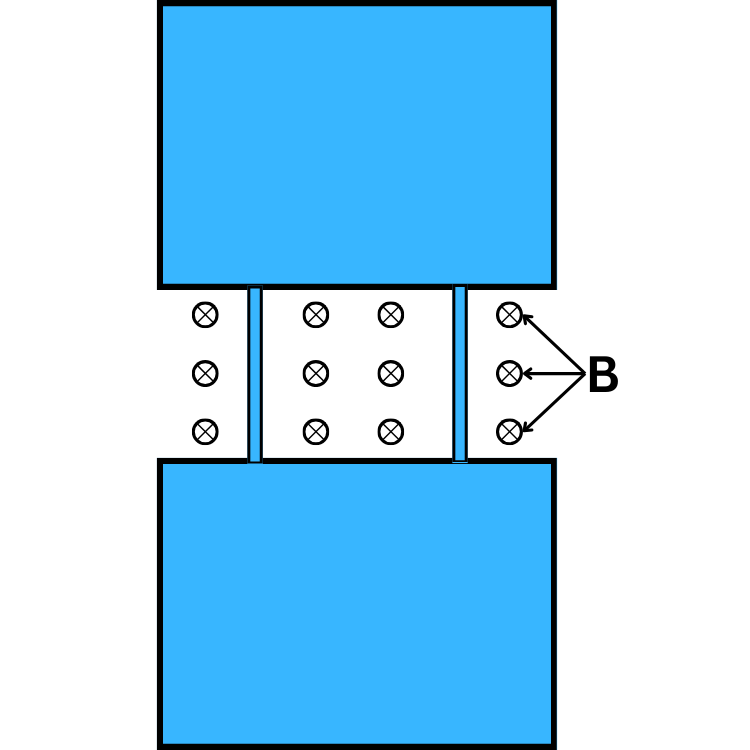}
     \caption{\textbf{Schematic of a Dayem loop qubit.} It involved two superconducting thin-film electrodes connected by two parallel nanowires, which are assumed identical. The electrodes form a coplanar capacitor $C$. The magnetic field $B$ is applied perpendicular to the device, as shown by the circles with "x" symbols in them.}
     \label{fig:device}
 \end{figure}

Initially, we analyze this system using a simplified generic model of the qubit CPR (termed "Likharev CPR" from Ref.\cite{RevModPhys.51.101}). Two main conclusions follow from this analysis. (1) If the critical current of the wire is too large (in practice, this is usually the case) and the corresponding frequency of the resulting transmon is too high, then it is always possible to reduce the qubit frequency by increasing the wire inductance by making the nanowires longer, since the inductance is linearly proportional to the length. (2) As the magnetic field is increased, the phase shift between the wires increases proportionally. This leads to a reduction of the linear inductance of the device while the nonlinear inductance remains unchanged. Therefore, the relative anharmonicity increases with magnetic field as a Lorentzian function and diverges as the critical current of the device goes to zero.

We also study a more realistic CPR, previously predicted by microscopic theory calculations \cite{PhysRevLett.103.087003}. We demonstrate that such zero-temperature CPR can be approximated by a simple power-law formula, which we use to analyze the anharmonicity of Dayem loop qubits. For the higher-power nonlinearity, the main conclusions are as follows. (1) The qubit frequency can still be adjusted by changing the wire length, since the linear term in the kinetic inductance is proportional to the wire length in any scenario. (2) At zero magnetic field, the nonlinear cubic term is absent, so the anharmonicity is probably too small for practical qubits. (3) At non-zero magnetic fields, a cubic term appears in the CPR of the device, and then the absolute anharmonicity can be increased by increasing the magnetic field. The relative anharmonicity shows an even stronger increase with the magnetic field, compared with the simplified cubic-nonlinearity CPR.

\section{Structure of the Dayem loop qubit}

A schematic of the Dayem loop qubit is shown in Fig.~\ref{fig:device}. For simplicity, we always assume that the nanowires are identical to each other. Two nanowires are connected, in parallel, to larger electrodes (blue rectangles). These electrodes play the role of antennas (if the device is measured within a 3D microwave cavity) and also play the role of a shunting capacitor $C$, which impacts the energy levels of the qubit as well as its anharmonicity and the Coulomb charging energy $E_c=e^2/2C$ ($e$ is the absolute value of the single-electron charge). Generally speaking, a transmon qubit can be viewed as an anharmonic oscillator, in which the oscillations arise from the back-and-forth flow of superconducting condensate electrons between the electrodes through the nanowires, acting as weak links between the electrodes. The analogy between the anharmonic oscillator and the transmon qubit is further explored in Appendix~\ref{apdx:transmon}. The design discussed here is similar to the traditional transmon qubit design \cite{transmon}, but the nanowires provide the weak links. We envision that nanowires would be on the order of 100 nm in length and $\sim$~10 nm in diameter, and the electrodes would be on the order of 100-500 $\mu$m in lateral dimension, and the thickness would be, say, $100$ nm, as in Ref.\cite{PhysRevB.82.134518}. The critical current of each nanowire might be of the order of 1 $\mu$A, as in, e.g., DNA-based nanowires \cite{hopkins-2005}.

The device is placed in a magnetic field perpendicular to the device plane (Fig.~\ref{fig:device}). This magnetic field creates a phase difference between the wires and a persistent bias current in the wires. The phase difference of wire "2" (the right wire) is related to the phase difference of wire "1" as $\phi_2=\phi_1+2\pi b$, assuming that there are no trapped vortices in the device \cite{sun-2025}. This phase shift is created and defined by the Meissner currents in the electrodes \cite{hopkins-2005}. This formula assumes that the phase gradients in the electrodes (antennas) do not depend on the current in the wires but only depend on the Meissner currents in the antennas themselves. In this formula, $b$ is the normalized magnetic field defined as $b=B/\Delta B$, where $B$ is the applied magnetic field and $\Delta B$ is the periodicity of the superconducting quantum interference device (SQUID), created by the two parallel nanowires.  

Below, we will consider two models for this device. The first model, simplified, is based on Ginzburg-Landau (GL) perturbation theory, and the second, more realistic model will be based on the zero-temperature current-phase relationship (CPR) obtained by solving Usadel equations and reported previously in Ref.\cite{PhysRevLett.103.087003}. 

\section{Simplified Dayem Loop qubit model}

\subsection{Current-Phase Relationship}

In this section, we consider a simplified nanowire current-phase relationship (CPR) derived from the GL theory in Appendix~\ref{apdx:GL}. Such a CPR will be referred to as the "Likharev CPR"\cite{RevModPhys.51.101} and is given as:

\begin{equation}
I(\phi)
=
\frac{3}{2}\,
I_c
\left[
\frac{\phi}{\phi_c}-\frac{1}{3}\frac{\phi^3}{\phi_c^3}
\right]
\end{equation}
Here $\phi_c$ is the critical phase difference at which the supercurrent $I$ is at its maximum. Note that this critical phase is proportional to the ratio of the nanowire length $L$ and the superconducting coherence length $\xi$ as $\phi_c=L/(\xi\sqrt{3})$. Here $I_c$ is the maximum possible current, which is achieved if $\phi=\phi_c$. It is called the critical current since at this current, superconductivity breaks down. 

To simplify notations, we rewrite the CPR as 

\begin{equation}\label{eq:cubic_cpr}
    \frac{I(\phi)}{I_0} = \frac{\phi}{\phi_{0}} - \left[\frac{\phi}{\phi_{0}}\right]^{3}
\end{equation}

Here $\phi$ is the phase difference of the condensate wave function taken between the ends of the wire, $I$ is the supercurrent in the wire, $\phi_0=\sqrt3\phi_c$, and $I_0=(3\sqrt3/2)I_c$.   According to the approximate CPR given above, $I=0$ if $\phi=\phi_0$. The maximum of this CPR occurs at $\phi/\phi_0=1/\sqrt{3}$ and the maximum supercurrent at that point is $2I_0/3\sqrt{3}$, so this value is the actual critical current if one measures just one nanowire. 

Now, we introduce a perpendicular magnetic field bias $B$, which generates a phase gradient in the electrodes that is linear with $B$. Physically speaking, the magnetic field creates Meissner screening currents, which induce the phase gradients in the direction in which the vector potential is zero \cite{hopkins-2005}. Now, let us say, the period of the superconducting quantum interference device (SQUID) formed by the two parallel wires is $\Delta B$. Then the phase shift between the wires equals $2\pi$ when $B=\Delta B$. Let us introduce a normalized magnetic field $b = B/\Delta B$. If the phase difference on wire-1 is $\phi_1$, then the phase difference on wire-2 is $\phi_2 = \phi_1 + 2\pi b$. Now, define a global phase difference between the electrodes as $\phi = (\phi_1 + \phi_2)/2$, the total supercurrent of the device is:

\begin{equation}\label{eq:toy_curr}
    \frac{I(\phi)}{I_{0}} = \frac{\phi_1}{\phi_{0}} - \left[\frac{\phi_1}{\phi_{0}}\right]^{3}+
    \frac{\phi_2}{\phi_{0}} - \left[\frac{\phi_2}{\phi_{0}}\right]^{3}
\end{equation}

or

\begin{equation}\label{eq:cubic_tot_curr}
    \frac{I(\phi)}{I_{0}} = \frac{2\phi}{\phi_0} - \frac{(\phi - \pi b)^{3} + (\phi + \pi b)^{3}}{\phi_0^{3}}
\end{equation}

Here, we assume that there are no trapped vortices between the wires or in the electrodes.

Expanding out the CPR obtains the following equation:

\begin{equation}
    \frac{I(\phi)}{I_0} = \left(\frac{2(\phi_0^2 - 3(\pi b)^2)}{\phi_0^3}\right)\phi - \left(\frac{2}{\phi_0^3}\right)\phi^3
    \label{cpreq3}
\end{equation}

\begin{figure}[t]
    \centering
    \includegraphics[width=1\linewidth]{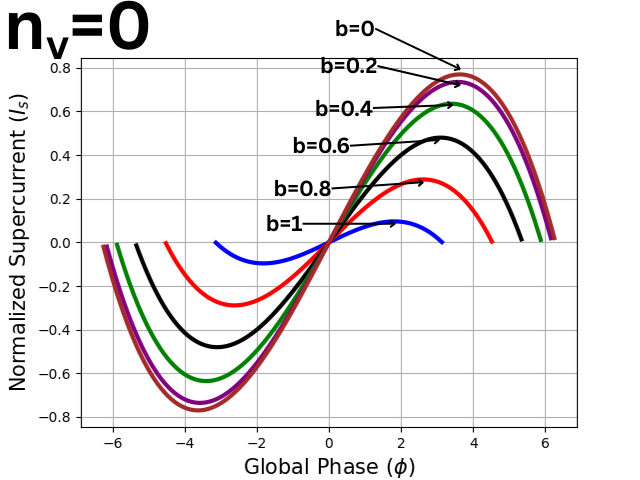}
    \includegraphics[width=1\linewidth]{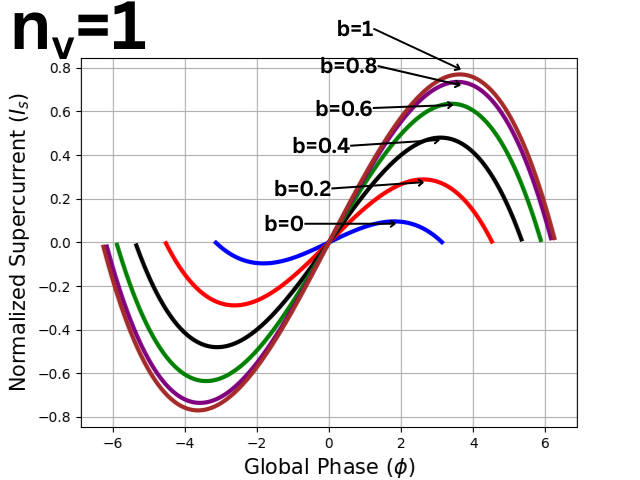}
    \caption{\textbf{Evolution of the total supercurrent versus magnetic field for $n_v=0$ and $n_v=1$.} For sufficiently high values of $b-n_v$, the local maxima and minima (i.e., kinetic inductance divergence points) are suppressed towards zero supercurrent.}
    \label{fig:toy_cpr_b}
\end{figure}

It is imperative to note that applying a non-zero phase shift between the nanowires (see Eq.~\ref{cpreq3}) suppresses the linear term in the CPR, but not the cubic term. This means that by increasing the phase shift between the wires, which equals $\phi_{shift}=2\pi b$, one can make the cubic term more and more significant in comparison to the linear term. This is a key finding that provides a solution to making nanowire qubits. 

Another interesting observation is that if the critical current of each wire is too large, one can make the critical current of the qubit smaller by increasing the phase shift $\phi_{shift}$. Let us show this explicitly. Let us now write CPR in the form 
\begin{equation}\label{cpr6}I/I_0=A\phi-D\phi^3=A\phi\left(1-\frac{D}{A}\phi^2\right)
\end{equation}
According to Eq.~\ref{cpreq3}, the coefficients defining the quadratic and the quartic terms are:
\begin{equation}
    A=A(b)=\frac{2(\phi_0^2 - 3\pi^2b^2)}{\phi_0^3}
    \label{EqA}
\end{equation}
\begin{equation}
    D=\frac{2}{\phi_0^3}
    \label{EqD}
\end{equation}

Then the location of the maximum of the current is defined by zero derivative, which occurs at $\phi^2=\phi_{max}^2=A/3D$. For example, in zero field the maximum is at $\phi^2=\phi_0^2/3=\phi_c^2$, as expected. Then the maximum current (i.e., the critical current) of the Dayem loop is $I_{max}=AI_0\phi_{max}(1-(D/A)\phi_{max}^2)=(2/3)AI_0\sqrt{A/3D}$.  Thus, the critical current is $I_{max}=(2/3\sqrt3)I_0A^{3/2}D^{-1/2}=(2/3\sqrt3\sqrt2)I_0(2/\phi_0-6\pi^2b^2/\phi_0^3)^{3/2}(\phi_0^3)^{1/2}$. From this one observes that the critical current of the SQUID goes to zero if the magnetic field satisfies $1-3\pi^2b^2/\phi_0^2=0$ or $b=\phi_0/\sqrt3\pi=\phi_c/\pi$, where $\phi_c$ is the critical phase of one nanowire, as introduced above. This makes sense since $\phi_{shift}=2\pi b$, so if $b=\phi_c/\pi$ then the phase bias on each wire, at zero external bias current, equals its critical phase. Therefore, at the field $b=\phi_c/\pi$, the Meissner current in the wires equals their critical current. 

The key property of the Dayem loop is that when $A(b)$ approaches zero, $D$ remains constant. So the interference effect suppresses the critical current, but it does not suppress the nonlinearity.  Such a result will be crucial to realizing a qubit with a sufficiently high anharmonicity. 

It should be emphasized that if the current in each wire remains, at all times, a few percent lower than its critical current, then the phase slip rate is exponentially low if the temperature is near zero. Thus, the device's vorticity remains constant in time, even as the magnetic field is tuned to increase the CPR's nonlinearity.

\begin{figure}[t]
    \centering
    \includegraphics[width=1\linewidth]{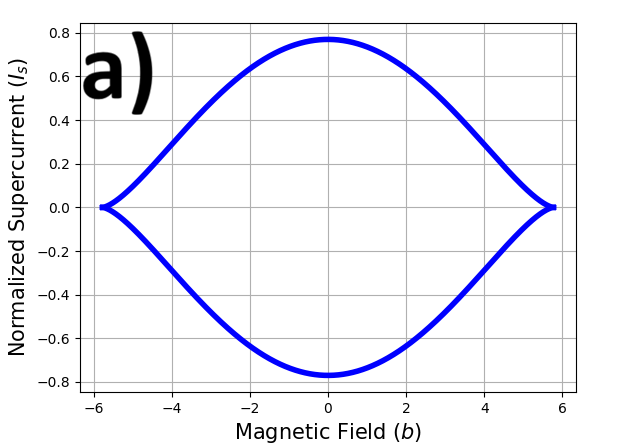}
    \includegraphics[width=1\linewidth]{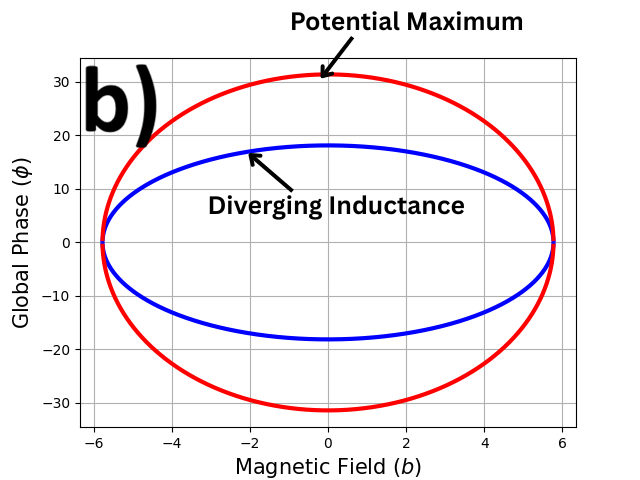}
    \caption{(a) $I_c(b)$ curve for $\phi_0 = 10\pi$. (b) Plots of $(\phi, b)$ with $\phi_0 = 10\pi$, where the inductance diverges (blue), and the potential reaches its maximum (red). Note that the potential energy maximum (red curve) occurs at a larger global phase difference as compared to the inductance divergence phase difference (blue curve). }
    \label{fig:I_v_phi}
\end{figure}

So far, we have assumed that there are no trapped vortices between the nanowires. Let us briefly consider what would happen if some number of vortices, $n_v$, is present in the loop formed by the two wires. Each vortex generates a total phase shift of $2\pi$, so the total phase shift becomes $\phi_{shift}=2\pi(b-n_v)$. (Note that each vortex in the SQUID loop contributes a phase bias of $\pi$ per wire. It is assumed that the electrodes act macroscopically, so the phase bias on them is negligible if a few vortices are present in the SQUID loop.) So the CPR becomes:

\begin{equation}
    \frac{I(\phi)}{I_0} = \left(\frac{2(\phi_0^2 - 3\pi^2(b-n_v)^2)}{\phi_0^3}\right)\phi - \left(\frac{2}{\phi_0^3}\right)\phi^3
    \label{CPREQ3}
\end{equation}

\begin{figure*}[t]
    \centering
    \includegraphics[width=0.48\linewidth]{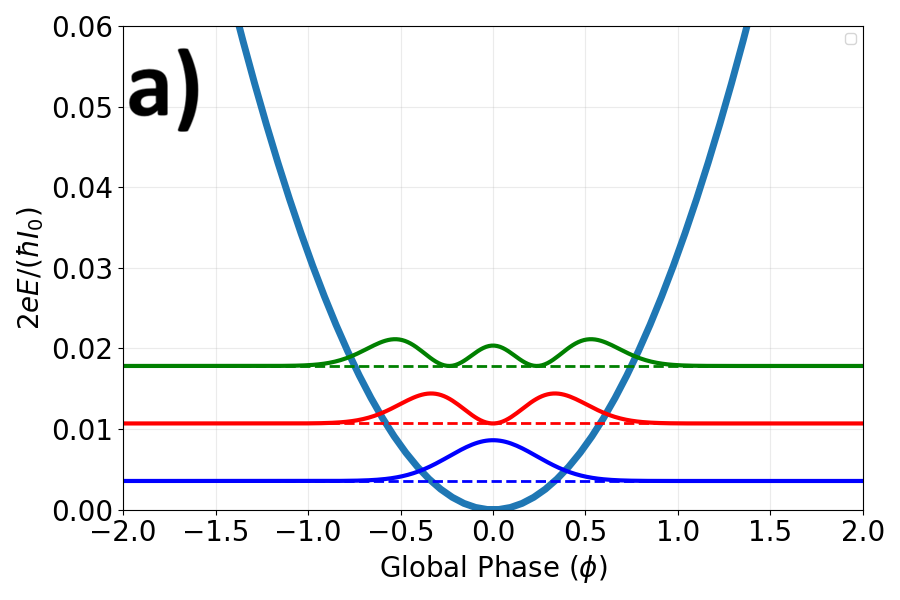}
    \includegraphics[width=0.48\linewidth]{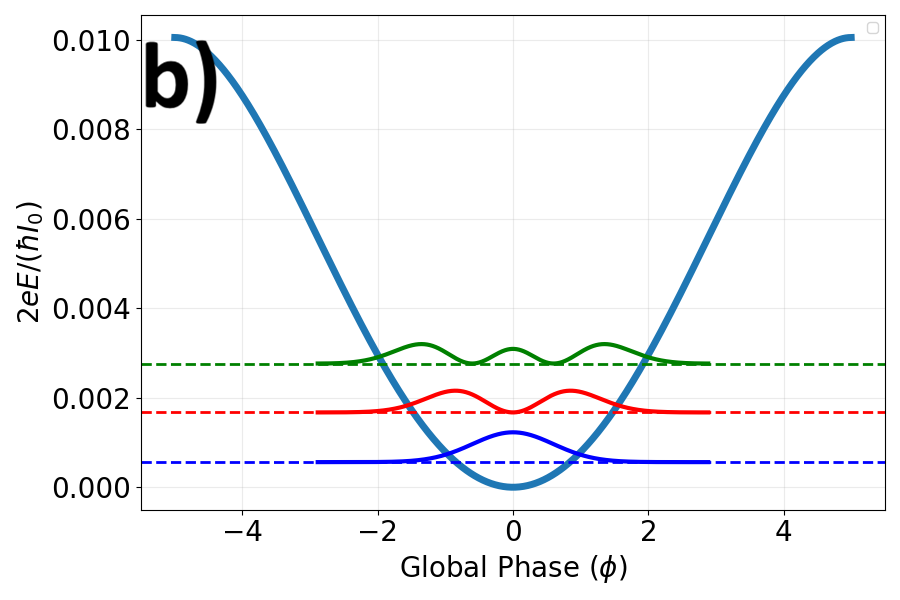}
    \caption{\textbf{Increasing anharmonicity of a Dayem loop qubit with a cubic order CPR via magnetic-field tuning.} Shown is the potential energy together with the lowest three eigenstates: ground (blue), first excited (red), and second excited (green). Solid curves represent the eigenfunctions, while dashed horizontal lines indicate the corresponding eigenenergies. The vertical axis shows the normalized energy $2eE/(\hbar I_0)$, with $n_v=0$, $I_0 = 1~\mu\mathrm{A}$, corresponding to an energy normalization factor $\hbar I_0/2e=497~\mathrm{GHz}$. The qubit parameters are $E_c/h \approx 0.3~\mathrm{GHz}$ and $\phi_0 = 10\pi$. The qubit excitation energy depends on the magnetic field. Note, all the presented energies below have units of $\hbar I_0/(2e)$. (a) \textit{Zero magnetic field ($b = 0$).} The maximum of the potential is $U_{max}=15.7$. The energy levels are $E_0$, $E_1$, and $E_2$. The ground state is $E_0 = 3.566 \times 10^{-3}$. The qubit excitation energy is $E_{01}=E_1-E_0 = 7.13\times10^{-3}$. Thus, the qubit excitation energy is $E_{01}=3.54$GHz. The second excitation energy gap is $E_{12}=E_2-E_1=E_{01}+\alpha$, where $\alpha=-2.427\times 10^{-6}$ is the absolute anharmonicity. This yields a small relative anharmonicity $\alpha_r =(E_{12}-E_{01})/E_{01}=-1+(E_{01}+\alpha)/E_{01}=\alpha/E_{01}=-0.00034$, which is $-0.034\%$. The small anharmonicity arises because the eigenfunctions are localized far below the potential maximum, where the potential is nearly harmonic, resulting in an almost linear energy spectrum. 
    (b) \textit{Finite magnetic field ($b = 5.7$).} The applied magnetic field pushes down the maximum of the potential, and the wave functions expand and become closer to the maximum. Thus, the effect of the non-linear terms of the potential is enhanced. The maximum of the potential energy is $U_{max}=1\times 10^{-2}$. The ground state energy is $E_0 = 5.6145 \times 10^{-4}$, the first excitation energy $E_{01} = E_1 - E_0 = 1.111\times 10^{-3}$, and the second one is  $E_{12} = E_2 - E_1 = 1.093\times 10^{-3}$. Therefore, the absolute anharmonicity is $\alpha = E_{12} - E_{01} = -1.8\times 10^{-5}$. This results in a significantly increased relative anharmonicity of $\alpha_r = (E_{12} - E_{01})/E_{01} = -0.016$ which is $-1.6\%$.}
    \label{fig:potentials}
\end{figure*}

This equation (Eq.~\ref{CPREQ3}) would be a representation of the Little-Parks effect if $n_v$ is allowed to adjust to minimize the quantity $|b-n_v|$. Yet, at near zero temperature, at which qubits normally operate, $n_v$ cannot change because the phase slip rate is controlled by the Arrhenius activation exponent, which is extremely small at low temperatures. The protection against phase slips occurs because the phase slips in superconducting wires have normal cores, and it costs significant condensation energy to suppress superconductivity even in a small segment of a nanowire.

This dependence of the total supercurrent on the normalized magnetic field is illustrated in Fig.~\ref{fig:toy_cpr_b} for $n_v=0$ and $n_v=1$. We observe that the maximum of the CPR shifts to a lower current as the magnetic field is increased. We also observe that it is possible to shift the local maximum of the total current function, $I(\phi)$, close to zero if the vorticity is changed. This is an important characteristic since for transmon qubit applications, the critical current has to be larger but comparable to the current introduced by just one energy quantum in the resonator LC-circuit, which constitutes the qubit. 

Also, according to Eq.~\ref{CPREQ3}, the critical current can be suppressed not only by changing the magnetic field but also by introducing vorticity into the loop. This can be achieved by driving the magnetic field to sufficiently high values. Since $n_v$ does not change at low temperature, the possibility of having different vorticity values $n_v$ constitutes a possibility to design programmable qubits if such a Dayem loop is used as the nonlinear inductive element of the qubit.

An example plot of the critical current versus magnetic field is shown in Fig.~\ref{fig:I_v_phi}a. There we find that the critical current can be pushed all the way to zero if the magnetic field is increased, assuming that phase slips do not occur in the wire and the vorticity, $n_v$, does not change.  That means that the nonlinear inductance regime can be pushed to as low a current as necessary with an external magnetic field. This is the key to achieving a transmon qubit functionality with a two-nanowire qubit, because in a transmon qubit, the current produced by just one microwave photon, absorbed by the qubit, has to be enough to change the slope of the CPR by at least a few percent.   


\subsection{Increasing anharmonicity with magnetic field in a simple model of a Dayem loop qubit}

To find the energy levels and the eigenstates of our proposed Dayem loop qubit, one needs to write its Hamiltonian. We need to choose its kinetic and potential energies, to build an analogy with a harmonic oscillator of slight anharmonicity. We will assume that the Coulomb charging energy of the device acts as its "kinetic energy" while the energy of the condensate in the nanowires, which is controlled by the phase difference $\phi$, acts as its "potential energy". Let us first construct the quantum operator of the kinetic energy. Classically, the kinetic energy of an electrical system is given by the Coulomb energy $T = Q^2/(2C)$ where $Q$ is the charge in the system and $C$ is the capacitance between the electrodes (antennas) connected to the nanowires. This term is always quadratic in $Q$. Note that the charge is proportional to the time derivative of the phase $\phi$, which is considered as the "coordinate" of the oscillator. This is because the voltage is $V=Q/C$ and the time derivative of the phase (coordinate) is $d\phi/dt=2eV/\hbar=2eQ/C\hbar$.

Similarly, the quantum mechanical "kinetic energy" operator for superconducting quantum circuits is $\hat{T} = \hat{Q}^2/(2C)$ where $\hat{Q} = 2e\hat{N}$. Here, $\hat{N}$ is the number operator that returns the number of pairs of electrons in the superconductor condensate, in excess of the charge neutrality point. In the phase basis, the number operator can be rewritten as $\hat{N} = i\partial_\phi$. Therefore, defining the Coulomb charging energy for one electron as $E_c = e^2/(2C)$, the kinetic energy term is $-4E_c \partial^2_\phi$. Here $e$ is the charge of one electron.

Moreover, the potential energy corresponding to any current-phase relationship (CPR) can be calculated as $U(\phi) = \hbar/(2e)\int_0^\phi  I_s(\phi') \, d\phi'$. Integrating Eq.~\ref{CPREQ3} one obtains the potential energy $U$ as $U(\phi)$ as:

\begin{equation}\label{eq:potential}
    U(\phi)= \left(\frac{\hbar I_0}{2e}\right)\left(\frac{1}{2}A\phi^2 -\frac{1}{4}D \phi^4\right)
\end{equation}

where $A$ and $D$ have been defined and discussed above. The maximum of this function occurs at $\phi=\sqrt{A/D}$ and the corresponding energy barrier is $U_{max}=\frac{\hbar I_0}{2e}\frac{A^2}{4D}=\frac{\hbar I_0}{2e}(\phi_0/2)(1-3\pi^2b^2/\phi_0^2)^2$. 

The potential energy function represents an approximately harmonic oscillator with a slight anharmonicity given by the fourth power (quartic) term. As was mentioned before, the introduction of a non-zero phase shift between the nanowires suppresses the quadratic term coefficient $A=A(b)=2/\phi_0-6\pi^2b^2/\phi_0^3$ while keeping the quartic term constant with the magnetic field. Such a result is again emphasized to be imperative towards realizing a sufficiently high anharmonicity in nanowire-SQUID transmons.

The phase difference $\phi$, also viewed as the "position" or "coordinate" of the oscillator, is described by the corresponding quantum mechanical operator $\hat{\phi}$. 
 
 The Hamiltonian is $\hat{H}=\hat{T}+\hat{U}$ is:

\begin{equation}\label{eq:cubic_Hamiltonian}
    \hat{H} = -4E_c\partial^2_\phi + \left(\frac{\hbar I_0}{2e}\right)\left(\frac{1}{2}A\phi^2 -\frac{1}{4}D \phi^4\right)
\end{equation} 

For this slightly anharmonic Hamiltonian, we first find the qubit excitation frequency $f_0$, in zeroth order in
$D$. (Here and always, $D=2/\phi_0^3=2/(3\sqrt3\phi_c^3)$, the coefficient in front of the quartic term, is assumed to be a small parameter.) 
The so-called plasma frequency of the qubit, $\omega$, is controlled by the quadratic term as 
\begin{equation}\label{eq:omega0}
\hbar\omega=
2\sqrt{\frac{E_c \hbar I_0A}{e}}
\end{equation}

(For comparison, note that for a traditional transmon qubit, this well-known formula would be $\omega=\sqrt{8E_cE_J}$, see Appendix~\ref{apdx:transmon}) Here, the parameter $A$ is dependent on the interference effect between the two wires and thus it is a function of the magnetic field. Note that for practical applications, the transmon plasma frequency should be in the range between 1 and 10 GHz, meaning that $\omega$ is in the range between 6x$10^9$ s$^{-1}$ and 6x$10^{10}$ s$^{-1}$. This is achieved by changing $E_c$ and $A$. The Coulomb charging energy $E_c$ is controlled by the geometry of the antennas, and $A$ is controlled by the length of the nanowires as well as by the applied magnetic field. If the critical current of the wire is too large (since it is difficult to make homogeneous nanowires with a critical current below 100 nA, which is usually needed for transmons with convenient and usable parameters), the magnetic field can be used to reduce $f_0$ by reducing $A$. 

The correspondence between $E_J$ and the coefficient in front of the quadratic term is $E_J \leftrightarrow (\frac{\hbar I_0}{2e})A$. Thus, zero-point fluctuations are

\begin{equation}
   \phi_{zpf}^4= \frac{2E_c}{A}\frac{2e}{\hbar I_0}=\frac{4e E_c\phi_0^3}{\hbar I_0(2\phi_0^2 - 6\pi^2 b^2)}
\end{equation}

As is derived in the Appendix~\ref{apdx:zpf}, the anharmonicity is proportional to the product of zero point fluctuations to the power four and the coefficient in front of the quartic term. In other words, the anharmonicity of a transmon qubit is $\alpha=-12K_4\phi_{zpf}^4$ where $K_4$ is the coefficient in front of the quartic term. In the considered case of the Dayem loop qubit, the coefficient $K_4$ is $K_4=(\frac{\hbar I_0}{2e}) \frac{D}{4}$. Then, its anharmonicity is calculated to be:
\begin{equation}
    \alpha=-12K_4\phi_{zpf}^4=-12\frac{(D/4)}{(A/2)}E_C=-\frac{6E_c}{\phi_0^2-3\pi^2b^2}
\end{equation}

Note that if $b=0$ then $\alpha=-6E_C/\phi_0^2$, where $\phi_0$ is a critical phase at which the superfluid density in a wire is suppressed, and the supercurrent becomes zero. 

We solve the Hamiltonian (Eq.~\ref{eq:cubic_Hamiltonian}) numerically using Python with boundary conditions $\small \psi(\pm  \sqrt{\phi_0^2/3 - \pi^2b^2}) = 0$. The physical meaning of the boundary conditions is that at the phase difference where the supercurrent goes to zero, we assume there is an infinite height potential wall. Later, we will verify that such a boundary condition has little to no effect on the wavefunctions because the wavefunctions are strongly localized at $\phi=0$.

The resulting evolution of the potential landscape, together with its associated eigenvalues and eigenfunctions as a function of the applied magnetic field, is shown in Fig.~\ref{fig:potentials}. As the magnetic field increases, the eigenfunctions shift closer to the top of the potential well, enhancing the influence of the nonlinear components of the potential on the energy spectrum. This increased nonlinearity leads to a corresponding enhancement of the anharmonicity. The energy is normalized by the constant $\hbar I_0/(2e)$, where $I_0 = 3\sqrt{3} I_c/2$ and $I_c$ denotes the critical current of each nanowire. We note that the metastable states remain localized near $\phi = 0$. Overall, these results demonstrate that the relative anharmonicity can be strongly enhanced by the application of a magnetic field.

To verify the accuracy of our numerically calculated metastable states, we utilize two techniques. First, we computed the energy levels using the Wentzel–Kramers–Brillouin (WKB) semiclassical approximation (Appendix~\ref{apdx:WKB}), finding good quantitative agreement with the numerical spectrum. 
Secondly, we also moved the infinite potential wells by about 10 percent, either closer to zero or away from zero. We observed a negligible change in the relative anharmonicity of about $1\%$. This result is because the wavefunctions are localized very close to the potential minimum, and approach zero probability exponentially as they approach the infinite potential energy wall. Therefore, the chosen boundary condition has no significant effect on the wavefunctions and the numerically calculated metastable states are accurate. 

\begin{figure}[t]
    \centering
    \includegraphics[width=1\linewidth]{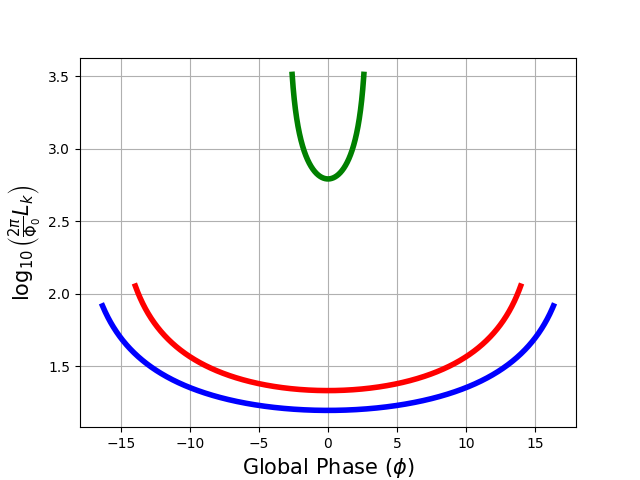}
    \caption{\textbf{Base-10 logarithm of the kinetic inductance after normalization by $\Phi_0/2\pi$ for a Dayem Loop qubit with a cubic CPR.} Here, we consider a device with $\phi_0 = 10\pi$ at magnetic fields of $b=0$ (blue), $b=3$ (red), $b=5.7$ (green).}
    \label{fig:inductance}
\end{figure}

\subsection{Inductance}

To achieve a high anharmonicity, the kinetic inductance of the device has to have a sufficient nonlinearity. Here, we derive the conditions necessary for the inductance to diverge. The objective is to push the nonlinear inductance regime towards sufficiently low supercurrents, e.g., similar to the transmon's operating currents. The kinetic inductance $L_k$ is  computed as:
\begin{multline}\label{eq:toy_inductance}
     L_k = \left(\frac{\Phi_0}{2\pi}\right)\left(\frac{dI}{d\phi}\right)^{-1} = \left(\frac{\Phi_0}{2\pi}\right)\frac{1}{I_0(A-3D\phi^2)}\\
     =\left(\frac{\Phi_0}{2\pi}\right)\frac{\phi_0^3}{I_0(2\phi_0^2 - 3(\phi  - \pi b)^2 - 3(\phi + \pi b)^2)} 
\end{multline}

The evolution of the log (base 10) of the kinetic inductance after normalization by $\Phi_0/2\pi$ with respect to the applied magnetic field is shown in Fig.~\ref{fig:inductance}. From Eq.~\ref{eq:toy_inductance}, it follows that the inductance diverges if:

\begin{equation}\label{eq:div_ind}
    \frac{\phi^2}{\phi_{\max}^2} + \frac{b^2}{b_{\max}^2} = 1
\end{equation}

Where $\phi_{\max} = \phi_0/\sqrt{3}=\phi_c$ and $b_{\max} = \phi_0/(\pi\sqrt{3})$.
A plot of Eq.~\ref{eq:div_ind} is shown in Fig.~\ref{fig:I_v_phi}b along with the curve corresponding to the maximum of the potential. Note that these curves do not align. The inductance divergence ellipses stay within the maximum potential ellipse. Thus, the device wave function might be able to sample the divergence region, which is the key factor allowing this nanowire-SQUID qubit to achieve higher anharmonicity compared to single-wire devices. Indeed, according to Eq.~\ref{eq:div_ind}, increasing $b$ will reduce, as low as necessary, the global phase at which the inductance diverges, as $\phi_{div}=\sqrt{\frac{A}{3D}}=\sqrt{\frac{\phi_0^2 - 3 \pi^2 b^2}{3}}=\sqrt{\phi_c^2 - \pi^2 b^2}$. This means that even if a single wire has a low anharmonicity, the SQUID will have a higher anharmonicity if the magnetic field is tuned. 

\subsection{Anharmonicity}

A high anharmonicity is necessary to reduce a many-level quantum system to an effective qubit.  The absolute and relative anharmonicity is defined as:

\begin{equation}
    \alpha = E_{12} - E_{01}, \;\;\; \alpha_r = \alpha/E_{01} 
\end{equation}

Where, $E_{12} = E_2 - E_1$ and $E_{01} = E_1 - E_0$. Next, consider the our weakly anharmonic Hamiltonian $\hat{H} = -4E_c\partial_\phi^2 +  (\hbar I_0/(2e))( A\phi^2/2 - D \phi^4/4)$.  The coefficients $A$ and $D$ are defined in Eq.~\ref{EqA} and Eq.~\ref{EqD}. We can apply time-independent perturbation theory to approximate the eigenenergies up to first order. Keeping the energy level $j$ arbitrary ($j$ is a non-negative integer), we can obtain the following approximation (See Appendix~\ref{apdx:perturbation} for more details):

\begin{multline}
    E_m \approx 2\sqrt{\frac{2E_c \hbar I_0 (\phi_0^2 - 3\pi^2 [b-n_v]^2)}{\phi_0^3e}}\left(j + \frac{1}{2}\right) \\\
    - \frac{2E_c}{\phi_0^2 - 3\pi^2[b-n_v]^2}(6j^2 + 6j + 3)
\end{multline}

The resulting expressions for the absolute and relative anharmonicity are:

\begin{align}
   \alpha &= -\frac{6E_c}{\phi_0^2 - 3\pi^2[b-n_v]^2} \\
   \alpha_r &= -\frac{3\phi_0^{3/2}}{2(\phi_0^2 - 3\pi^2[b-n_v]^2)^{3/2}}\sqrt{\frac{2e E_c}{\hbar I_0}}
\end{align}

Here, as always, $n_v$ is the number of fluxons trapped between the two nanowires. This result demonstrates that applying a phase shift, with either an applied magnetic field or by introducing vortices between the two nanowires, will increase anharmonicity. In particular, as the total phase shift between the wires increases, the absolute anharmonicity increases and the plasma frequency decreases. Both of these effects, therefore, increase the relative anharmonicity of the device. 

Such an effect is qualitatively different than the single nanowire qubit, where the anharmonicity is only dominated by the length of the nanowire and its critical current. We also observe that the excitation energy ($E_{01}=E_1-E_0$) is proportional to $\sqrt{I_0A}$, therefore, if $E_{01}$ is too large due to the critical current of the nanowires being too high, one can always reduce it by applying an appropriate magnetic field, since $A$ is reduced with magnetic field.

\section{Realistic Nanowire CPR Models}

\begin{figure*}[t]
    \centering
        \includegraphics[width=0.32\linewidth]{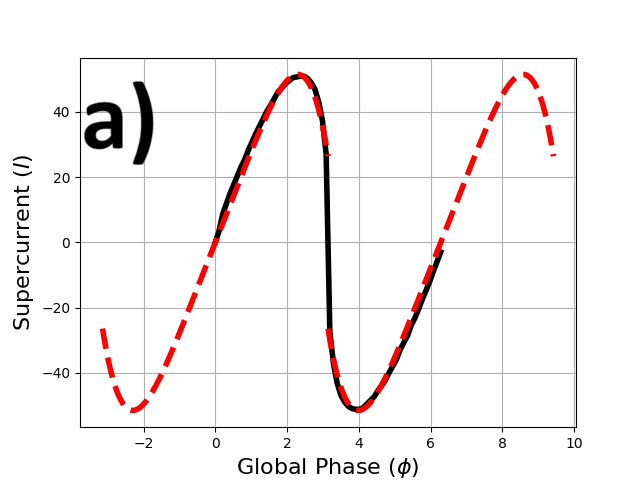}
        \includegraphics[width=0.32\linewidth]{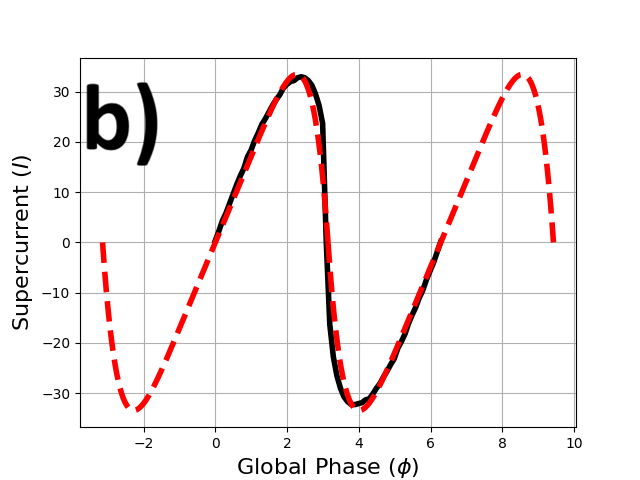}
        \includegraphics[width=0.32\linewidth]{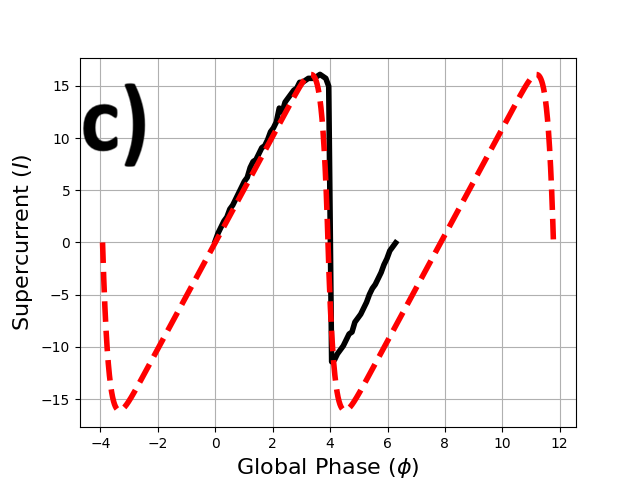}
                \caption{\textbf{Power-law fits to zero-temperature nanowire current--phase relationships.} The y axis (Supercurrent $I$) is in $\mu \rm A$. Black curves show current--phase relationships calculated from the Usadel equations in Ref.~\cite{PhysRevLett.103.087003} at $T=0.15$K with a nanowire width of $45$nm. Red curves show fits to the generalized power-law model of Eq.~\ref{eq:true_cpr_single}.     Across a wide range of parameters and CPR shapes, the model reproduces the numerically calculated curves with high accuracy, capturing both the extended linear regime at small phase bias and the sharp onset of nonlinearity near the critical current. (a) Here, the length of the nanowire is $15$nm. The corresponding power law fit had the following parameters: $m=2$, $a=0.7$, $I_0 = 88\mu \rm A$, $\phi_0=\pi$. (b) Here, the length of the nanowire is $45$nm. The corresponding power law fit had the following parameters: $m=3$, $a=1$, $I_0 = 54\mu \rm A$,  $\phi_0=\pi$. (c) Here, the length of the nanowire is $240$nm. The corresponding power law fit had the following parameters: $m=9$, $a=1$, $I_0 = 20\mu \rm A$, $\phi_0=\pi$. These results justify the use of the generalized $(2m+1)$ power-law CPR in the perturbative analysis that follows.}
    \label{fig:fits}
\end{figure*}

\subsection{Zero-temperature current-phase relationship (CPR) of superconducting nanowires}

In the previous sections, we studied a simplified model in which the nonlinear correction to the nanowire current-phase relationship (CPR) takes a cubic form. This approximation is valid at temperatures near the critical temperature or, potentially, for extremely short nanowires. However, transmon qubits operate near absolute zero temperature, where the CPR can deviate significantly from this cubic form. A more realistic zero-temperature model is therefore required to obtain reliable estimates of the proposed qubit performance.

Microscopic calculations based on the Usadel equations \cite{RevModPhys.76.411} have previously been carried out by Ref.\cite{PhysRevLett.103.087003}. We digitize their reported CPR curves and find that they can be modeled with high accuracy using a simple power-law form, as described below.

While Eq.~\ref{eq:cubic_cpr} accurately captures the CPR at high operating temperatures, the low-temperature CPR curves reported in Ref.\cite{PhysRevLett.103.087003} show a qualitatively different behavior. In particular, the CPR becomes increasingly linear at low phase bias as the temperature is reduced, leading to the onset of nonlinearity shifting closer to the critical current. To account for these deviations, we introduce a generalized power-law model that fits both the results of Ref.\cite{RevModPhys.51.101} and those of Ref.\cite{PhysRevLett.103.087003}:

\begin{equation}
\label{eq:true_cpr_single}
\frac{I(\phi)}{I_0} = \frac{\phi}{\phi_{0}} - a\left(\frac{\phi}{\phi_{0}}\right)^{2m+1}
\end{equation}

Here, $a$ is a dimensionless constant which defines the strength of the nonlinearity, and $m$ is a positive integer. 

This CPR gives zero current ($I=0$) if $\phi=\phi_{zero}=\phi_0a^{-1/2m}$. And the maximum of the current occurs at $\phi=\phi_{max}=\phi_0/(a(2m+1))^{1/2m}$. It is interesting to calculate the difference:
\begin{equation}
\phi_{\text{zero}} - \phi_{\text{max}}
=
\phi_0\, a^{-1/(2m)} \Big[\, a^{1/m} - (2m+1)^{-1/(2m)} \,\Big]
\end{equation}
An interesting observation is that as wires get longer and $m$ gets larger, this difference approaches zero, although it is always a positive number. Thus, for longer wires, we get the expected result that the maximum of the CPR occurs close to the point where the supercurrent drops to zero.

For example, if $m=2$ then $\phi_{zero}=\phi_0a^{-1/4}$.  Also, if $a=1$ then $I=0$ at $\phi_{zero}=\phi_0$ at any $m$. Also, for $m=2$, $\phi_{max}=\phi_0/(5a)^{1/4}$, which is less than $\phi_{zero}$ as expected.

A practical justification for the power-law CPR given above is obtained by directly comparing it to the CPR curves calculated using the microscopic theory and reported in Ref.~\cite{PhysRevLett.103.087003}. We find an excellent agreement, as is demonstrated in Fig.~\ref{fig:fits}. 

If the non-linear term is absent, then the CPR is exactly linear, and it is impossible to make a qubit. Such a case has been analyzed previously in relation to memory elements ~\cite{sun-2025}. But qubits are possible for any $m>0$, such as the Likharev CPR discussed above, $m=1$, since the nonlinear correction is cubic. Two features of this model are noted. In general, the nonlinear correction is presented as $2m+1$, to emphasize that the CPR remains an odd function of $\phi$. Second, as $m$ increases, the CPR nonlinearity shifts closer to the critical phase.  The cubic CPR is recovered as the special case $a=1$ and $m=1$, or, as will be discussed below, if two wires are connected in parallel, forming a SQUID.

\begin{figure*}[t]
    \centering
    \includegraphics[width=0.48\linewidth]{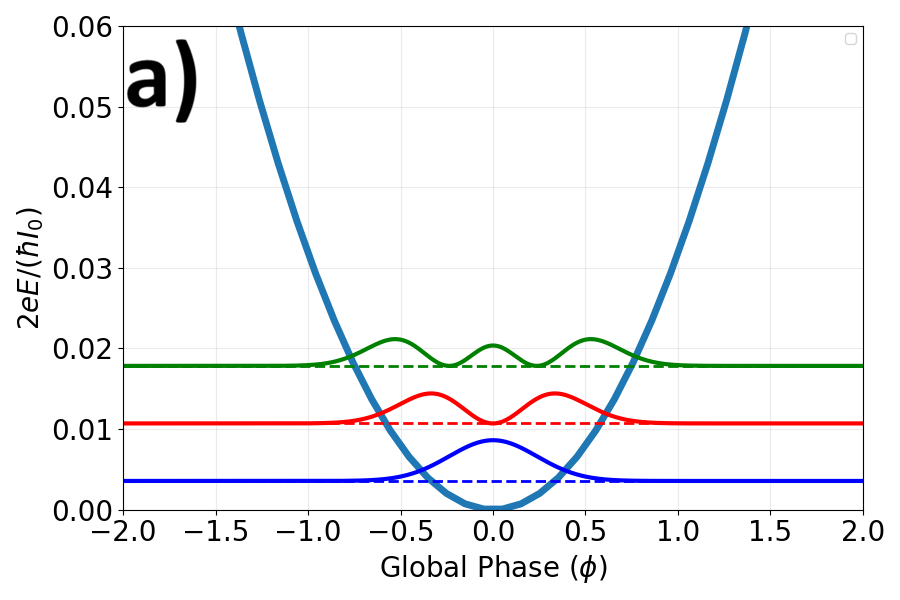}
    \includegraphics[width=0.48\linewidth]{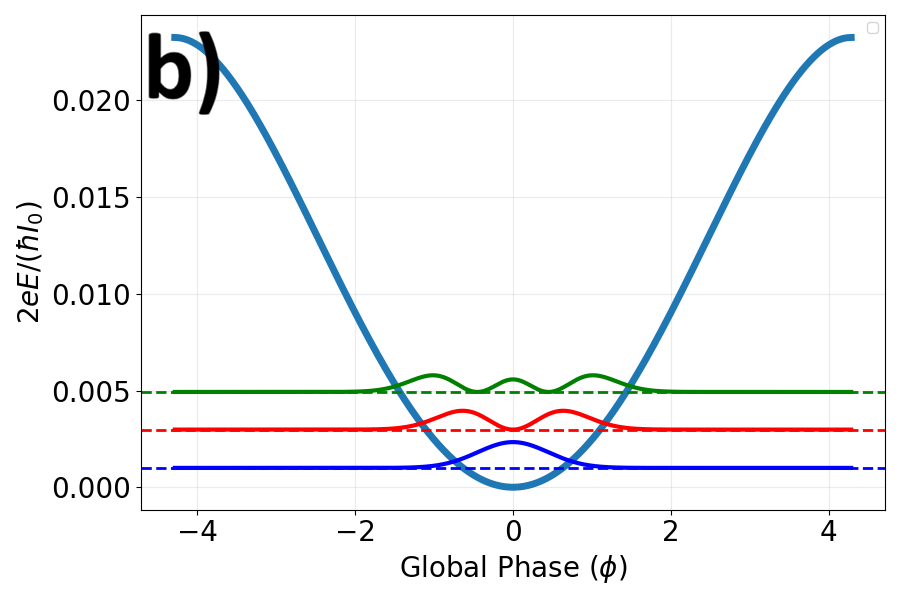}
    \caption{\textbf{Wave functions and anharmonicity of a Dayem loop qubit with a fifth-order CPR. Magnetic-field effect.} Shown is the potential energy together with the lowest three eigenstates: ground (blue), first excited (red), and second excited (green). Solid curves represent the eigenfunctions, while dashed horizontal lines indicate the corresponding eigenenergies. The vertical axis shows the normalized energy $2eE/(\hbar I_0)$, with $n_v=0$, $I_0 = 1~\mu\mathrm{A}$, corresponding to an energy normalization factor $\hbar I_0/2e=497~\mathrm{GHz}$. The qubit parameters are $E_c/h \approx 0.3~\mathrm{GHz}$ and $\phi_0 = 10\pi$. The qubit excitation energy depends on the magnetic field. Note, all the presented energies below have units of $\hbar I_0/(2e)$. Here, $a=1$. (a) \textit{Zero magnetic field ($b = 0$).} In the perturbative picture, at $b=0$, the potential reduces to a simple harmonic oscillator with zero anharmonicity. The energies are therefore $\hbar\omega(j + 1/2)$ where $j$ is an integer and $\hbar\omega = 4\sqrt{\hbar I_0 E_c / (2e\phi_0)} = 3.6196 \times 10^{-24}$, corresponding to $\omega/2\pi = 5.5$GHz.
    (b) \textit{Finite magnetic field ($b = 6.55$).} Similar to the cubic CPR case, the magnetic field pushes the wavefunctions closer to the maximum of the potential, thereby increasing the impact of the non-linearity. The maximum of the potential energy is $U_{max}=2.32\times 10^{-2}$. The ground state energy is $E_0 = 9.989 \times 10^{-4}$, the first excitation energy $E_{01} = E_1 - E_0 = 1.981\times 10^{-3}$, and the second one is  $E_{12} = E_2 - E_1 = 1.945\times 10^{-3}$. Therefore, the absolute anharmonicity is $\alpha = E_{12} - E_{01} = -3.6\times 10^{-5}$. This results in a significantly increased relative anharmonicity of $\alpha_r = (E_{12} - E_{01})/E_{01} = -0.018$ which is $-1.8\%$.}
    \label{fig:5th_potentials}
\end{figure*}

\subsection{Restoration of the cubic term in CPR due to quantum interference between two nanowires}

We now extend our power-law CPR model to the Dayem loop qubit by introducing a phase shift $\pi(b-n_v)$ between the wires, analogous to the treatment used for the cubic CPR. Here, $b$ denotes the applied perpendicular magnetic field (normalized), and $n_v$ is the integer number of vortices trapped between the nanowires. The total supercurrent of the device is then:

\begin{equation}
\label{eq:true_cpr_double}
\small
\frac{I(\phi)}{I_0} =
\frac{2\phi}{\phi_0}
- a
\frac{
(\phi - \pi[b-n_v])^{2m+1}
+
(\phi + \pi[b-n_v])^{2m+1}
}{\phi_0^{2m+1}}
\end{equation}

In what follows, we will assume $n_v=0$. Our main finding is that even if $m>1$, we can still get a desirable cubic nonlinearity if two wires are connected in parallel and a magnetic field is amplified between them. The magnetic field creates a phase shift $\phi_{shift}=2\pi b$. The cubic term of the SQUID CPR occurs as follows. For one wire, the CPR is $i_1=\delta_1-a\delta_1^{2m+1}$, where we use the normalized current $i_1=I_1/I_0$ and the normalized phase difference on the wire, $\delta_1=\phi_1/\phi_0$. The expression for wire 2 is the same except that $\delta_2=\phi_2/\phi_0=(\phi_1+\phi_{shift})/\phi_0=\delta_1+\delta_{shift}$. And $a$ is a constant defining the amplitude of the nonlinear term. The global normalized phase difference is $\delta=(\delta_1+\delta_2)/2$. With these notations, the SQUID normalized CPR is:

\begin{multline}
    i=(\delta-\delta_{shift}/2)+(\delta+\delta_{shift}/2)  \\
    - a(\delta-\delta_{shift}/2)^{2m+1}-a(\delta+\delta_{shift}/2)^{2m+1}    
\end{multline}

The first, linear, term of the expression is
\[ i_{linear}(\delta) \approx \left[ 2 - 2a(2m+1)\left(\frac{\delta_{\text{shift}}}{2}\right)^{2m} \right]\delta  \]
Apparently, this term can be reduced by the phase shift $\delta_{shift}=\phi_{shift}/\phi_0=2\pi b/\phi_0$. In fact, it can be reduced to any low value, even to zero. In general, the linear term controls the frequency of the qubit. Thus, the conclusion is that the frequency can be controlled in a wide range. Here, it is assumed that the vorticity is zero and that there are no phase slips (which change the vorticity) during the qubit operation. This assumption is reasonable since to create a phase slip in a nanowire, one needs to create a normal region, which costs the condensation energy, which is usually quite large for a homogeneous nanowire.

The cubic correction term of the general CPR is
\[ i_{cubic}(\delta)  = - 2a \binom{2m+1}{3} \left(\frac{\delta_{\text{shift}}}{2}\right)^{2m-2}\delta^3
\]
It is linear at zero magnetic field, but it increases in magnitude if a magnetic field is applied, which generates the phase shift $\delta_{shift}$. Thus, the calculations done in the previous sections, based on the cubic CPR, are relevant in general since the cubic term can be restored due to quantum interference between the two nanowires.

In the following sections, we perform a perturbative analysis of a few examples and compute the resulting relative anharmonicity for (i) the fifth-order CPR ($m=2$), (ii) the seventh-order CPR ($m=3$), and (iii) the general case of arbitrary $m$. We demonstrate that even nanowires with highly linear CPRs can support appreciable anharmonicity, provided that a sufficient phase shift is induced between the two wires. This establishes a practical route toward building qubits from otherwise weakly nonlinear nanowire elements.

\subsection{CPR with fifth order nonlinearity}
In this section, we analyze the total supercurrent and the resulting anharmonicity for a symmetric SQUID with a fifth-order ($m=2$) current--phase relationship (CPR). The total current is
\begin{equation}\label{eq:fifth_tot_curr}
    \frac{I(\phi)}{I_0} 
    = \frac{2\phi}{\phi_0} 
    - a\frac{(\phi - \pi[b-n_v])^{5} + (\phi + \pi[b-n_v])^{5}}{\phi_0^{5}}
\end{equation}

To simplify the discussion, we assume $n_v=0$. Then
\begin{equation}\label{eq:expanded_fifth_cpr}
\frac{I(\phi)}{I_0}
=
\left(
\frac{2}{\phi_0}
-
\frac{10a\pi^4 b^4}{\phi_0^5}
\right)\phi
-
\frac{20a\pi^2 b^2}{\phi_0^5}\phi^3
-
\frac{2a}{\phi_0^5}\phi^5
\end{equation}

Interestingly, a term cubic in $\phi$ appears in the current, due to quantum interference between the wires. It is clear that the linear term is now controlled by the magnetic field $b$ and can be made as small as necessary to achieve the right qubit frequency. The cubic term actually gets larger in magnitude if the magnetic field is increased. The fifth power term does not depend on the magnetic field. The linear term goes to zero if \begin{equation}
b_{div} = \frac{\phi_0}{\pi (5a)^{1/4}}
\end{equation} 
This corresponds exactly to the condition that the Meissner current in the electrodes induces a phase shift on each nanowire such that the current in each nanowire equals its maximum possible supercurrent (i.e., its critical current). 

We restrict our analysis to highly localized wavefunctions, $|\phi| \ll \phi_c$, and expand the CPR in powers of $\phi$ up to third order. This approximation yields an effective quartic potential of the form
\begin{equation}
\label{effective}
\small
     U(\phi) \approx \frac{\hbar I_0}{2e}
     \left[
     \frac{\phi_0^4 - 5a\pi^4 b^4}{\phi_0^5}\phi^2  
     - \frac{5a\pi^2 b^2}{\phi_0^5}\phi^4
     \right]
\end{equation}

Two key features emerge from this expression. First, the Meissner phase shift for the nanowires (assuming $n_v=0$), is $\pi b$, suppresses the quadratic term through the factor $\phi_0^4 - 5a(\pi b)^4$. Second, and in contrast to the cubic CPR case ($m=1$), the quartic term is explicitly increasing by the phase shift, scaling as $b^2$. By increasing $b^2$, one simultaneously decreases the harmonic term and enhances the non-linear correction, thereby amplifying the overall nonlinearity of the device. One can compare this to the case when the CPR of each wire is cubic. In that case, the quartic term is independent of the applied phase difference. In the case considered here, when the CPR of each wire has a fifth-power nonlinearity, the phase shift directly strengthens the nonlinear component of the potential.

For the vortex-free arrangement ($n_v=0$) the effective potential energy of the Dayem loop qubit is shown in Eq.\ref{effective}. Even though the CPR of each wire does not have a cubic correction, the entire device has a cubic correction in the CPR, which correspond to a quartic correct in the potential energy.

The plasma frequency of such a qubit can be found using Appendix A and Appendix E and equals

\begin{equation}
\hbar\omega
=
2\sqrt{
\frac{2E_C \hbar I_0}{e}
\left(
\frac{\phi_0^4-5a\pi^4b^4}{\phi_0^5}
\right)
}.
\end{equation}

Alternatively, if $E_c$ is expressed in GHz units, then the qubit excitation frequency $f=\omega/(2\pi)$ can be written as

\begin{equation}
f
=
\frac{1}{2\pi}\sqrt{
\frac{16E_C[GHz] I_0}
{e(5a)^{1/4}\,b_{\mathrm{div}}}
\left[
1-
\left(
\frac{b}{b_{\mathrm{div}}}
\right)^4
\right]
}.
\end{equation}
During qubit operations the magnetic field has to be lower than its value at which the current in a nanowire reaches its critical value, i.e., $b<b_{div}=\frac{\phi_0}{\pi (5a)^{1/4}}$. At $b=b_{div}$ superconductivity is suppressed.

The anharmonicity can be obtained using first-order time-independent perturbation theory. Matching the coefficients of the effective potential (Eq.\ref{effective}) to the standard quartic oscillator form given in Appendix A and Appendix~\ref{apdx:perturbation} yields a closed-form approximation for the absolute and relative anharmonicity as a function of magnetic field:

\begin{align}\label{eq:rel_anharmonicity_fifth}
    \alpha &= -\frac{60 E_c a\pi^2b^2}{\phi_0^4 - 5a\pi^4b^4} 
\end{align}

\begin{equation}
    \alpha_{r} 
    =\frac{\alpha}{\hbar\omega}= -\sqrt{\frac{2eE_c}{\hbar I_0}}
    \frac{15a\pi^2 b^2\phi_0^{5/2}}
    {\left(\phi_0^{4} - 5a\pi^4 b^4\right)^{3/2}}\end{equation}   D

Here, we present an example set of parameters that allow a high qubit frequency (W-band, between 75 and 110 GHz) and a good strong anharmonicity ($\alpha_r=0.01$). An example set of parameters could be $I_0=10 \mu A$, $a=1$, $\phi_0=3\pi$, $E_c=2.5GHz$,  $b = 1.68$. (Note that $b_{div}=2$ in this case.)

The corresponding qubit frequency is $103$ GHz and the anharmonicity is $\alpha=-0.419E_c=-1.05 GHz$. Thus the relative anharmonicity is about $1\%$. The anharmonicity can be easily increased. For example, if $b-1.85$ then the qubit frequency is $\omega/2\pi=76GHz$ and the anharmonicity is $\alpha=-0.928E_c=-2.32GHz$. So the relative anharmonicty is about 3\%.

This example further emphasizes the importance of the phase shift between nanowires. In particular, when $b=0$, then in the perturbative picture, the quartic contribution vanishes and the relative anharmonicity becomes zero. Therefore, a finite phase shift (finite $b$) is necessary to achieve a nonzero anharmonicity in the fifth-order CPR.

Suppose the desired qubit frequency is just a few GHz, as is customary in modern quantum computers. This value can also be achieved as the following example shows. Consider an example $a=1$, $n_v=0$, $\phi_0=10\pi$, $E_c/h \approx 0.3~\mathrm{GHz}$, and $I_0 = 1~\mu\mathrm{A}$. In this case, the quadratic term goes to zero at $b=(\phi_0/\pi)/(5a)^{1/4}=6.68$. The qubit excitation energy depends on the magnetic field (see below). Such an example is shown in Fig.~\ref{fig:5th_potentials}. There, all the presented energies are normalized by $\hbar I_0/(2e)=497~\mathrm{GHz}$, and plotted on the graphs are the potential energy together with the lowest three eigenstates: ground (blue), first excited (red), and second excited (green). Solid curves represent the eigenfunctions, while dashed horizontal lines indicate the corresponding eigenenergies. The vertical axis shows the normalized energy $2eE/(\hbar I_0)$. 
 
 The qubit excitation energy depends on the magnetic field. The case of zero magnetic field ($b = 0$) is shown in Fig.~\ref{fig:5th_potentials}a. According to the Appendix~\ref{apdx:zpf}, the frequency of such a qubit is $\omega^2=4K_1K_2$, where $K_1=4E_C/\hbar^2$ and $K_2=\frac{\hbar I_0}{2e}(1/\phi_0 - 5a(\pi b)^4/\phi_0^5)$. So $\omega^2=\frac{8E_C I_0}{e\hbar}(1/\phi_0 - 5a(\pi b)^4/\phi_0^5)$. Neglecting all terms beyond the cubic term in the CPR, and for $b = 0$, the potential reduces 
to that of a simple harmonic oscillator, implying vanishing anharmonicity 
($\alpha_r = 0$) in this approximation. The quantized energy levels are therefore 
approximately $\hbar\omega(j + 1/2)$ where $j$ is an integer and $\hbar\omega = 4\sqrt{\hbar I_0 E_c / (2e\phi_0)} = 5.776 \times 10^{-24}$J, corresponding to the qubit excitation frequency of 8.7 GHz and zero anharmonicity ($\alpha_r=0$) within this approximation. 

Next, if higher order corrections are taken into account, then the anharmonicity is (see the Appendix~\ref{apdx:zpf}(6)):
 $\alpha=-180K_6\phi_{zpf}^6=-180K_6(E_C/K_2)^{3/2}$. Let us calculate first zero point fluctuations of the phase for $b=0$ (see the Appendix~\ref{apdx:zpf}):
 \begin{equation}
\phi_{\mathrm{zpf}}^4 = \frac{\hbar^2}{4}\frac{K_1}{K_2}=\phi_0\frac{2eE_C}{\hbar I_0}
\label{eq:phi_zpf1}
\end{equation}

Substituting the numbers chosen above, one gets $\phi_{zpf}=0.2367$. Thus, the size of the wave function is much smaller than the critical phase, as it should be for the perturbation model to work well. Next, we evaluate the coefficient $K_6$ by integrating Eq.~\ref{eq:expanded_fifth_cpr} for $b=0$. Thus $K_6=\frac{a}{3\phi_0^5}\frac{\hbar I_0}{2e}=1.089\times10^{-8}\times 497$GHz. With this we obtain the anharmonicty $\alpha=-180\times (0.2367)^6 \times 1.089\times10^{-8}\times 497$GHz=171Hz,  which is very small. The relative anharmonicity in this case is $\alpha_r=$171Hz/$E_{01} = 171\mathrm{Hz}/(\hbar \omega)= 171\mathrm{Hz}/8.7\mathrm{GHz} = -1.97\times10^{-8}$, which is only $0.00000197\%$. This is too small for any practical qubit application.

A numerical Hamiltonian solver was employed to compute the relative anharmonicity. However, due to grid discretization effects, the solver is unable to reliably resolve extremely small anharmonicities. In particular, when $a = 0$, the Hamiltonian reduces to that of a quantum harmonic oscillator, for which the relative anharmonicity is exactly $\alpha_r = 0$. Nevertheless, the numerical solver yields a finite value, denoted $\alpha_{r,a=0}$, which is about an order of magnitude larger than the perturbative estimate of $10^{-8}$ obtained in the previous paragraph. Consequently, the relative anharmonicity computed for $a = 1$, denoted $\alpha_{r,a=1}$, is found to be of the same order as $\alpha_{r,a=0}$, indicating that it is dominated by numerical error. Although $\alpha_{r,a=0}$ can be reduced to approximately $10^{-7}$ by refining the discretization, $\alpha_{r,a=1}$ remains of comparable magnitude. This value therefore serves as an upper bound for $\alpha_{r,a=1}$. So it is concluded that in the case $b=0$ the numerical error dominates the computed result. In general, both the perturbative and numerical approaches consistently indicate that, at zero magnetic field, the relative anharmonicity is extremely small for the chosen set of parameters. Yet, as is discussed below, an appropriately chosen magnetic field can mitigate this problem and increase the anharmonicity to about 1\%, which is good enough for practical applications of the Dayem loop qubit. 
 
 An example numerical calculation result for a finite magnetic field ($b=6.55$) is given in Fig.~\ref{fig:5th_potentials}b. Similar to the cubic CPR case, the magnetic field pushes the maximum of the potential to the localization region of the wavefunction, thereby increasing the impact of the non-linearity. For the parameters given above, the normalization constant is as before $\hbar I_0/(2e)=497~\mathrm{GHz}$. The maximum of the potential energy is $U_{max}=2.32\times 10^{-2}(\hbar I_0/(2e))=11.5$GHz. The ground state energy is $E_0 = 9.989 \times 10^{-4}(\hbar I_0/(2e))$, the first excitation energy $E_{01} = E_1 - E_0 = 1.981\times 10^{-3}(\hbar I_0/(2e))=0.985$GHz, and the second one is  $E_{12} = E_2 - E_1 = 1.945\times 10^{-3}(\hbar I_0/(2e))=0.967$GHz. Therefore, the absolute anharmonicity is $\alpha = E_{12} - E_{01} = -3.6\times 10^{-5}(\hbar I_0/(2e))=0.018$GHz. This results in a practically acceptable relative anharmonicity of $\alpha_r = (E_{12} - E_{01})/E_{01} = -0.018$ which is $-1.8\%$.

 Another method is as follows. The relative anharmonicity, if a cubic term is present in the CPR and thus a quartic term is present in the potential energy, is $\alpha=-12K_4\phi_{zpf}^4$. The quartic factor is $K_4=\frac{\hbar I_0}{2e}\frac{5a\pi^2 b^2}{\phi_0^5}=6.9\times10^{-5}\times 497$GHz=34.4MHz. Now, the zero point fluctuations is given by $\phi_{zpf}^4 = (\hbar^2/4) (K_1/K_2) = (0.44559)^4$, the absolute anharmonicity can be calculated as $K_4 \times -12 \times (0.44559)^4 = -3.27 \times 10^{-6} \times 497$GHz $= -16.8$MHz. Thus, the relative anharmonicity is $-16.8\text{MHz}/967\text{MHz}=-0.017$ or $-1.7\%$, with good agreement with numerical simulations. 

  Yet one more evaluation method is given by Eq.~\ref{eq:rel_anharmonicity_fifth}. If we put the parameters chosen above in Eq.~\ref{eq:rel_anharmonicity_fifth}, we get a relative anharmonicity of -0.016, which is -1.6\%. Again, we observe good agreement with numerical results. 
   
More generally, this provides the first concrete illustration of a broader trend: although, at low temperatures, in longer nanowires $m>1$ and so the nanowire CPRs are highly nonlinear only very near the critical current, the introduction of a non-zero phase difference between the wires can drive the nonlinearity to lower currents by inducing a cubic nonlinearity. Also, the phase shift between the wires reduces the qubit frequency by reducing the total critical current. Since a phase shift can be induced by a small external magnetic field (typically a few Gauss, induced by an external small solenoid), such an effect enables homogeneous nanowire qubits, without requiring any interfaces like in gatemons, for example, or any external current bias.  

\subsection{Seventh Order CPR}

In this section, we extend our perturbative analysis on the symmetric SQUID towards nanowires exhibiting a seventh-order ($m=3$) current--phase relationship (CPR). The total current is

\begin{equation}\label{eq:seventh_tot_curr}
    \frac{I(\phi)}{I_0} = \frac{2\phi}{\phi_0} - a\frac{(\phi - \pi[b  - n_v])^{7} + (\phi + \pi[b- n_v])^{7}}{\phi_0^{7}}
\end{equation}

Similar to the fifth-order CPR section, we will restrict our analysis to highly localized wavefunctions, and expand this supercurrent out to cubic order in $\phi$. Integrating this CPR yields a quartic potential function:

\begin{equation}    
\small
     U(\phi) \approx \frac{\hbar I_0}{2e}\left[\frac{\phi_0^6 - 7a(\pi[b-n_v])^6}{\phi_0^7}\phi^2  - \frac{35a(\pi[b-n_v])^4}{2\phi_0^7}\phi^4\right]
\end{equation}

The structure of the potential energy for the seventh-order CPR parallels that of the fifth-order case, with a stronger dependence on the phase difference between the two nanowires. Suppose, for example, that there are no trapped vortices in the loop ($n_v=0$). Then, as the magnetic field increases, the quadratic term in the potential is suppressed through the factor $\phi_0^6 - 7a(\pi b)^6$, while the quartic term is increased proportionally to $b^4$. 
Such a result reflects the fact that even if the power of the nonlinearity in the CPR is 7, it is still possible to gain a cubic nonlinearity in the CPR and, correspondingly, a quartic nonlinearity in the potential energy.

Physically, this behavior implies that nanowires whose CPRs are approximately linear at low currents can still be driven into a strongly anharmonic regime by introducing a sufficiently large phase imbalance between the two nanowires. The higher power of $(\pi[b-n_v])$ appearing in the quartic term indicates that the onset of nonlinearity is more sensitive to magnetic fields in higher-order CPRs than in lower-order ones. As a result, devices governed by a seventh-order CPR require larger phase shifts to achieve appreciable anharmonicity.

Note, the maximum magnetic field that the device can support is equivalent to the magnetic field for which the quadratic term is exactly zero. Therefore, $b_{\max} = \phi_0/(\pi (7a)^{1/6})$. In the below examples, $\phi_0 = 10\pi$ and $a=1$, therefore $b_{\max} \approx 7.23$.

Using Appendix~\ref{apdx:perturbation}, we can calculate the relative anharmonicity of the device as:

\begin{equation}
    \alpha_{r} = -\sqrt{\frac{2eE_c}{\hbar I_0}}\frac{105a(\pi[b-n_v])^{4}\phi_0^{7/2}}{2(\phi_0^{6} - 7a(\pi[b-n_v])^{6})^{3/2}}
\end{equation}

Moreover, the plasma frequency is then:

\begin{equation}\label{eq:gen_freq}
\omega =
\sqrt{
8\frac{E_C I_0}{e\hbar}
\frac{
\phi_0^{6}
-7a\bigl[\pi(b-n_v)\bigr]^{6}
}
{\phi_0^{7}}
}.
\end{equation}

As an example, we will calculate the relative anharmonicity with the parameters $E_c/h \approx 0.3$GHz, $\phi_0 = 10\pi$, $I_0 = 1\times 10^{-6}$. At $b=0$, in the perturbative picture, the anharmonicity is zero since the potential function reduces to a simple harmonic oscillator with energies $\hbar\omega(j + 1/2)$ $j$ is an integer and $\hbar \omega = \sqrt{\hbar I_0 E_c/(2e\phi_0)} = 3.6196 \times 10^{-24}$J, corresponding to $\omega/2\pi = 5.5$GHz. 

Next, we apply $b = 7$ (which is less than $b_{\max} \approx 7.23$) to the system and assume $n_v=0$ (no trapped vortices). Then the ground state energy is $E_0 = 1.4912 \times 10^{-3}\frac{\hbar I_0}{2e}$, and the first excitation energy is $E_{01} = E_1 - E_0 = 2.9682 \times 10^{-3}\frac{\hbar I_0}{2e}$. The second excitation energy is $E_{12} = E_2 - E_1 = 2.938\times 10^{-3}\frac{\hbar I_0}{2e}$. Therefore, the absolute anharmonicity is $\alpha = E_{12} - E_{01} = -3.02 \times 10^{-5}\frac{\hbar I_0}{2e}$. Then, the relative anharmonicity is $\alpha_r = \alpha/E_{01} = -0.011$ or $-1.1\%$, which is sufficiently strong for transmon operations. Previously it was shown that transmons with 1\% anharmonicity are possible to operate\cite{purkayastha2026tunableanharmonicitysninasnanowire}.  

We note that, in contrast to the fifth-order CPR relative anharmonicity, a larger magnetic field was necessary to push the anharmonicity up to one percent. Such an effect is reflected in the fact that in the seventh-order CPR, the nonlinear regime is shifted closer to the critical current. Thus, a larger phase difference is required to drive the system to its nonlinear regime. Moreover, we continue to observe a general trend that nanowires with higher order nonlinearities in the CPRs can be used as nonlinear inductors for qubits, provided that the phase shift between the wires is made sufficient by the magnetic field.

\subsection{General Case}

In this section, we generalize our perturbative analysis towards symmetric SQUIDs with nanowires, demonstrating a current phase relationship on the order of $2m+1$, where $m$ is treated as an arbitrary integer. The total supercurrent of the device is:

\begin{multline}\label{eq:general_tot_curr}
    \frac{I(\phi)}{I_0} = \frac{2\phi}{\phi_0} \\
    - a\frac{(\phi - \pi[b  - n_v])^{2m+1} + (\phi + \pi[b- n_v])^{2m+1}}{\phi_0^{2m+1}}
\end{multline}

To fully expand Eq.~\ref{eq:general_tot_curr}, we employ the binomial theorem to expand the polynomials $(\phi \pm \pi[b- n_v])^{2m+1}$. Firstly, integrating this term to yield the potential energy, then collecting terms up to the quartic order, corresponds to the following:

\begin{multline}
    U(\phi) = \frac{\hbar I_0}{2e}\Bigg[\frac{\phi_0^{2m} - a(2m+1)(\pi[b-n_v])^{2m}}{\phi_0^{2m+1}}\phi^2 \\
    - \frac{a(4m^2-1)(m)(\pi[b-n_v])^{2(m-1)}}{6\phi_0^{2m+1}}\phi^4\Bigg] \\
    \iff \frac{\hbar I_0}{2e}\left[\frac{1}{2}A \phi^2 - \frac{1}{4}B \phi^4\right]
\end{multline}

The generalized plasma frequency of this device is then:

\begin{align}
    \hbar \omega &= 2\sqrt{\frac{E_c \hbar I_0A}{e}} \\
    &= \sqrt{\frac{8E_c \hbar I_0}{e}\frac{\phi_0^{2m} - a(2m+1)(\pi[b-n_v])^{2m}}{\phi_0^{2m+1}}}  
\end{align}

Here, this potential function emphasizes a general trend observed in highly linear CPRs with $m > 1$. In general:

\begin{equation*}
\text{Increasing $|b-n_v|$:}\quad
\text{quadratic term} \downarrow
\quad
\text{quartic term } \uparrow
\end{equation*}

Such a result has the effect of driving the system towards the non-linear regime as the magnetic field increases, and so the phase bias between the two nanowires increases. Moreover, the effect of the phase bias is increased as the CPRs approach a highly linear behavior for small $\phi$. This reflects the fact that as $m$ increases, the non-linear regime is pushed towards currents near the critical current. Therefore, a larger phase bias is required to push the total system relative anharmonicity towards a few percent. Then, using Appendix~\ref{apdx:perturbation}, we arrive at the formula for the relative anharmonicity of an $2m+1$-th order CPR.

\begin{equation}
\small
    \alpha_{r} = -\sqrt{\frac{2eE_c}{\hbar I_0}}\frac{a(2m+1)(m)(2m-1)(\pi[b-n_v])^{2(m-1)}\phi_0^{m+1/2}}{2(\phi_0^{2m} - a(2m+1)(\pi[b-n_v])^{2m})^{3/2}}
\end{equation}

The general expressions derived above unify the behavior observed in the cubic, fifth, and seventh order CPRs. As the order $m$ increases, the CPR becomes increasingly linear near $\phi=0$, while the nonlinear corrections are pushed closer to the critical current. However, the presence of a phase difference between the two nanowires can overcome this linearity by simultaneously suppressing the quadratic term and amplifying the quartic term in the effective potential. 

In general, a larger $\phi_0$ decreases the relative anharmonicity of the entire system. Therefore, shorter nanowires are favorable towards building qubits.  

In general, increasing the magnetic field enhances the anharmonicity for all $m > 1$, with the sensitivity to the phase shift growing rapidly with CPR order.
Importantly, these results demonstrate that even nanowires with highly linear zero-temperature CPRs can function as qubits by simply applying a phase shift between the wires' magnetic design alone.

\section{Discussion}

In this section, we discuss the fabrication and operation feasibility of our proposed Dayem loop qubit with current state-of-the-art techniques. 

We first discuss the feasibility of operating a Dayem loop qubit at a high plasma frequency. As previously mentioned, qubits operating in the W-band (75--110 GHz) may provide advantages for suppressing the influence of thermally excited as well as excess quasiparticle fluctuations, thereby decreasing the decoherence rate. We now evaluate the plasma frequency of the proposed Dayem loop qubit using realistic device parameters. As a reminder to the reader, the formula for the generalized plasma frequency was derived as:

\begin{equation}\label{eq:gen_freq}
\omega =
\sqrt{
\frac{8E_c I_0}{e\hbar}
\frac{
\phi_0^{2m}
-a(2m+1)\bigl[\pi(b-n_v)\bigr]^{2m}
}
{\phi_0^{2m+1}}
}.
\end{equation}

Moreover, the formula for the generalized relative anharmonicity was derived as:

\begin{equation}
\small
    \alpha_{r} = -\sqrt{\frac{2eE_c}{\hbar I_0}}\frac{a(2m+1)(m)(2m-1)(\pi[b-n_v])^{2(m-1)}\phi_0^{m+1/2}}{2(\phi_0^{2m} - a(2m+1)(\pi[b-n_v])^{2m})^{3/2}}
\end{equation}

For a target plasma frequency in the W-Band $\omega/2\pi \sim 75$ GHz, where decoherence effects due to quasi-particle fluctuations are mitigated, such a frequency scale can be achieved by optimizing the device parameters. In particular, such device parameters include the critical current $I_0$, the charging energy $E_c$, and the linearity of the nanowires' CPR (dictated by $m$). As an example, we calculate $\alpha_r$ and $\omega$ with the following device parameters: $I_0 = 1 \times 10^{-3}$ A, $n_v=0$, $a=1$, $E_c/h = 0.3$ GHz, $\phi_0 = 10\pi$, and $m=20$. Note that $m=20$ corresponds to a nanowire with a highly linear CPR, except very near the critical current. Then, we find that at $b=9.03$, the Dayem loop qubit achieves a plasma frequency of $\omega = 145.66$ GHz while maintaining a relative anharmonicity of $\alpha_r = 1\%$. Moreover, the induced supercurrent in each nanowire was calculated to be $88.8\%$ of its critical current. Therefore, we conclude that our Dayem loop qubit is capable of operating in the W-band frequency regime while maintaining a sufficiently high relative anharmonicity.

Next, we discuss the mitigation of magnetic field cross-talk. The Dayem Loop qubit relies on a perpendicular magnetic field to restore the cubic non-linearity of the qubit's Hamiltonian. But the required field is very small, of the order of 1 Gauss. We have previously tested that such a small magnetic field does not adversely affect the qubit\cite{PhysRevB.82.134518}. This is because at low fields vortices do not penetrate the structural components of the qubit, assuming that the applied field is less than the first critical magnetic field of its components. 

Similar challenges have been extensively considered in superconducting qubit platforms that utilize magnetic flux as a parameter, such as flux-tunable qubits \cite{yan-2016} and fluxonium qubits \cite{doi:10.1126/science.1175552}. In these systems, magnetic crosstalk is mitigated through techniques such as compensation pulses that correct for residual flux coupling between neighboring circuit elements \cite{neill-2018}, crosstalk calibration techniques \cite{PRXQuantum.2.040313}, and electromagnetic shielding \cite{malevannaya-2025}. If needed, such mitigation strategies can be implemented into the Dayem Loop qubit architecture.

Next, we discuss the suppression of phase slips in the Dayem loop qubit. Note that the Dayem loop qubit is designed to operate well below its critical temperature ($T_c$). However, due to the small diameter of the nanowires, phase slips, which represent vortices crossing the nanowires, might, in principle, contribute to the decoherence effects. Yet, a previous test\cite{PhysRevB.70.214506} showed that the phase slips are exponentially suppressed at low temperatures, even in extremely thin and narrow Dayem bridges. For example, one sample (B2\cite{PhysRevB.70.214506}) had a width of only 28 nm and a thickness of 3.5 nm. The extrapolated resistance was found as low as $10^{-8}$ Ohms at about 2.7 K. The corresponding typical current is 100 nA. The voltage is $10^{-15}$ V. Such parameters can then be converted to the phase slip rate, corresponding to a rate of phase slips of only 1 per second, which is much larger than the decoherence rate of modern transmons. Thus the contribution of phase slips is expected to be negligible. Moreover, in the same example above, the Dayem bridge is much thinner than its width. It still operates according to the established models. The parameters, such as the critical current, of the weak link depend on its cross section. 

The operation temperature of the proposed qubit should be much lower than the $T_c$ of the metal or the compound used to make nanowires. This way we ensure that all or almost all electrodes are in the superconducting condensate. Modern qubits are made of Al, and the operating temperature is below 100 mK. Yet, there are many different metals, alloys, or compounds with critical temperatures ten or even a hundred times higher than that of aluminum. Thus the operating temperature can, in principle, be made ten or a hundred times higher compared to the typical modern transmon qubits.

Next, we discuss the tunability of the relative anharmonicity of the Dayem loop qubit. According to Ref\cite{purkayastha2026tunableanharmonicitysninasnanowire}, a relative anharmonicity of about 1\% is sufficient. Moreover, our proposed device is flexible. In the section following Eq.32 we provide an example of qubit parameters leading to a qubit frequency of 103 GHz and a relative anharmonicity of about 1\%. The anharmonicity can be easily increased. For example, if $b$ is made somewhat higher, then the qubit frequency
decreases to $\omega/2\pi = 76$ GHz and the relative anharmonicity increases to 3\%. In the case analyzed in relation to Eq.32 the anharmonicity is proportional to $b^2$, which means it can be increased by increasing $b$. An example corresponding to a 3\% anharmonicity is given below Eq.32. In fact, anharmonicity diverges at $b=b_{div}$, according to Eq.32.

An interesting published prediction related to the qubit anharmonicity is that the nonlinear effects in the CPR of nanowires can be enhanced by fabricating bilayer
nanowires involving a sandwich structure of a superconductor and a normal metal\cite{vodolazov-2018}. The current phase relationship of such bilayer nanowires is strongly non-linear, although not sinusoidal, as demonstrated in\cite{vodolazov-2018}. Based on the general shape of the predicted CPR, it appears that the presence of a thin normal layer might reduce the degree of the nonlinear power law $m$.  Because the CPR exhibits a strong non-linearity, such a bilayer nanowire "sandwich" can be used as a qubit. An exact quantitative analysis of the SN sandwich nanowire CPR is beyond the scope of our current investigation.

Finally, we discuss the feasibility of fabricating such nanowires. Firstly, superconducting nanowires with dimensions comparable to those considered in this paper have been successfully fabricated, for example, using various suspended templates including nanobridges\cite{PhysRevB.70.214506} and molecular templating techniques\cite{hopkins-2005}, including DNA templating, which produce ultrathin nanowires with widths of only a few nanometers while maintaining lengths greater than their superconducting coherence length.  

In practice, fabrication imperfections such as local variations in wire width, thickness, crystallinity, or material composition introduce degrees of inhomogeneity. Such imperfections in superconducting nanowires have been extensively investigated experimentally and reviewed in the literature \cite{Bezryadin_book, Bezryadin-2000, koblischka-2021}. It was concluded that superconductivity remains strong with imperfections, even when the material is so disordered that it appears amorphous. Previous theoretical analysis of such nanowire superconducting devices has shown that moderate asymmetry between two parallel nanowires produces only a relatively small asymmetry in the critical current, and consequently, the effective non-linear inductance of such a device \cite{sun-2025}. We therefore expect that realistic fabrication imperfections will lead primarily to minor changes in device parameters rather than qualitatively altering the device's physical behavior. Also, the material can be made amorphous, while still showing excellent superconducting characteristics, as in Ref\cite{PhysRevB.70.214506}.

\section{Conclusion}

We proposed a qubit design involving two parallel supercondu    cting nanowires, which can be called "Dayem loop qubit" since it does not contain tunnel junctions or any interfaces between different materials. For a nanowire with a current-phase relationship (CPR) of cubic non-linearity, we find closed curves representing the inductance divergence points, which appear elliptical in the phase difference - magnetic field plane. We then applied perturbation theory to approximate the anharmonicity.
We present evidence that for longer wires, the current-phase relationship (CPR) has a power-law deviation from the linear behavior, in which the power becomes larger as the wire is made longer. This fact makes it more difficult to make qubits with longer nanowires. Yet, we demonstrate that this problem can be mitigated if two nanowires form a superconducting quantum interference device (SQUID). In this case, the interference between the wires leads to an increased nonlinearity by generating a cubic term in the current-phase relationship (CPR) of the device. This is true even if the nonlinearity of the CPR of each wire taken separately is of a higher than cubic order. In this case, we used first-order perturbation theory to approximate the anharmonicity of such nanowires. Then, we found that even though superconducting nanowires by themselves might have a low level of nonlinearity, which is apparent only very near the critical current, the interference effect between the two nanowires leads to a cubic nonlinearity restoration in the CPR of the entire device, thus enabling qubit operations. 

\appendix

\section{\label{apdx:transmon}Analogy Between a Generic Anharmonic Oscillator and Superconducting Qubits}

In this appendix, we establish the analogy between a generic weakly anharmonic oscillator and the superconducting transmon qubit. We then extend the same framework to the Dayem Loop transmon qubit with a non-sinusoidal current–phase relation (CPR).

\subsection{Generic Anharmonic Oscillator}

We consider a one-dimensional oscillator with Hamiltonian
\begin{equation}
H = \frac{\hat{p}^{2}}{2m_1} + \frac{1}{2}m_1\omega^{2}x^{2} + U_{\mathrm{anh}}(x)
\iff K_1 \hat{p}^2+K_2 x^{2}-K_4 x^{4}
\label{eq:H_generic}
\end{equation}
where $m_1$ is the mass of the particle and  \(U_{\mathrm{anh}}(x)=-K_4 x^4\) is a weak nonlinear correction to the harmonic potential. The generic coefficients are defined as $K_1=1/(2m_1)$ and $K_2=m_1\omega^2/2$.

The harmonic part defines the angular frequency 

\begin{equation}
\omega=2\sqrt{K_1K_2}.
\label{angfr}
\end{equation}
The harmonic part also defines the zero-point fluctuation amplitude
\begin{equation}
x_{\mathrm{zpf}}^{4} = \frac{\hbar^2}{4}\frac{K_1}{K_2}
= \left(\frac{\hbar}{2m_1\omega}\right)^2
\label{eq:xzpf}
\end{equation}

The anharmonicity of the spectrum is
\begin{equation}
\alpha = (E_2-E_1)-(E_1-E_0)=E_2-2E_1+E_0
\label{eq:alpha_def}
\end{equation}

To first order in perturbation theory,
\begin{equation}
\alpha =
\langle2|U_{\mathrm{anh}}(\hat x)|2\rangle
-2\langle1|U_{\mathrm{anh}}(\hat x)|1\rangle
+\langle0|U_{\mathrm{anh}}(\hat x)|0\rangle
\end{equation}

For a purely quartic perturbation \(U_{\mathrm{anh}}(x)=-K_4x^4\), this evaluates to
\begin{equation}
\alpha = -12K_4x_{\mathrm{zpf}}^{4}
= -3\hbar^2\frac{K_1K_4}{K_2}
\label{eq:alpha_generic}
\end{equation}

Thus the anharmonicity scales with the fourth power of the zero-point fluctuations.

\subsection{Transmon Qubit as a Weakly Anharmonic Oscillator}

The transmon Hamiltonian is
\begin{equation}
H_{\mathrm{tr}} = 4E_c N^2 - E_J \cos\phi
\label{eq:H_transmon}
\end{equation}
where \(\phi\) is the superconducting phase difference and \(N\) is the Cooper-pair number operator with
\begin{equation}
[\phi,N]=i, \qquad N=-i\frac{\partial}{\partial\phi}
\end{equation}

In the transmon regime \(E_J \gg E_c\), the Josephson potential can be expanded as
\begin{equation}
- E_J\cos\phi \approx -E_J
+ \frac{1}{2}E_J\phi^2
- \frac{E_J}{24}\phi^4 + \cdots 
\label{eq:cos_expansion}
\end{equation}

Identifying
\[
x \leftrightarrow \phi
\qquad 
p \leftrightarrow \hbar N
\qquad 
m_1 \leftrightarrow \hbar^2/(8E_c)
\]
the Hamiltonian maps onto Eq.~(\ref{eq:H_generic}) with
\[
K_1=\frac{4E_c}{\hbar^2}
\qquad 
K_2=\frac{E_J}{2}
\qquad 
K_4=\frac{E_J}{24}
\]

The quadratic term defines a harmonic oscillator with frequency
\begin{equation}
\omega = \frac{\sqrt{8E_cE_J}}{\hbar}
\end{equation}
and phase zero-point fluctuations
\begin{equation}
\phi_{\mathrm{zpf}}^{4}=\frac{2E_c}{E_J}
\label{eq:phi_zpf}
\end{equation}

Using Eq.~(\ref{eq:alpha_generic}), the transmon anharmonicity becomes
\begin{equation}
\alpha_{\mathrm{tr}}
= -12\left(\frac{E_J}{24}\right)\phi_{\mathrm{zpf}}^{4}
= -E_c
\label{eq:alpha_tr}
\end{equation}
which is the standard first-order result.

\subsection{Dayem Loop Qubit}

For a Dayem Loop qubit with a CPR of cubic order, the Hamiltonian can be written as
\begin{equation}
\begin{split}
\hat{H} =-4E_c \partial_\phi^2+U(\phi)= \\
=-4E_c \partial_\phi^2+K_2\phi^2+K_4\phi^4= \\
=
-4E_c \partial_\phi^2
+
\left(\frac{\hbar I_0}{2e}\right)
\left(
\frac{1}{2}A\phi^2
-
\frac{1}{4}D\phi^4
\right)
\label{eq:Hamiltonian}
\end{split}
\end{equation}

The harmonic part defines the angular frequency 

\begin{equation}
\omega=2\sqrt{K_1K_2}.
\label{angfr}
\end{equation}
For example, if the nonlinearity is cubic, then the coefficients are
\begin{equation}
A(b)=\frac{2}{\phi_0}-\frac{6\pi^2 b^2}{\phi_0^3}
\qquad
D=\frac{2}{\phi_0^3}
=\frac{2}{3\sqrt3\,\phi_c^3}
\end{equation}

Matching to Eq.~(\ref{eq:H_generic}) gives
\[
K_1=\frac{4E_c}{\hbar^2}
\qquad
K_2=\frac{\hbar I_0}{2e}
\left(
\frac{1}{\phi_0}-\frac{3\pi^2 b^2}{\phi_0^3}
\right)
\qquad
K_4=\frac{\hbar I_0}{2e}\frac{1}{2\phi_0^3}
\]

The zero-point phase fluctuations are therefore
\begin{equation}
\phi_{\mathrm{zpf}}^{4}
=
\frac{\hbar^2}{4}\frac{K_1}{K_2}
=
\frac{E_c}{(\hbar I_0/2e)}
\frac{\phi_0}{1-3\pi^2 b^2/\phi_0^2}
\label{eq:xzpf2}
\end{equation}

Using Eq.~(\ref{eq:alpha_generic}), the anharmonicity becomes
\begin{equation}
\alpha =
-12K_4\phi_{\mathrm{zpf}}^{4}
\label{eq:alpha_SQUID1a}
\end{equation}

or

\begin{equation}
\alpha =-12K_4\frac{\hbar^2}{4}\frac{K_1}{K_2}
\label{eq:alpha_SQUID1b}
\end{equation}

In particular, for a cubic nonlinearity in the CPR we get

\begin{equation}
\alpha =
-\frac{6E_c}{\phi_0^2-3\pi^2 b^2}
\label{eq:alpha_SQUID}
\end{equation}

If the nonlinear correction in the CPR is of the fifth order (i.e., m=2) then $K_1$ remains unchanged and 

\begin{equation}
\small
     U(\phi) =K_2\phi^2-K_4\phi^4= \frac{\hbar I_0}{2e}
     \left[
     \frac{\phi_0^4 - 5a\pi^4 b^4}{\phi_0^5}\phi^2  
     - \frac{5a\pi^2 b^2}{\phi_0^5}\phi^4
     \right]
\end{equation}
Thus 
\begin{equation}
K_2
=
\frac{\hbar I_0}{2e}
\frac{\phi_0^4-5a\pi^4b^4}{\phi_0^5}
=
\frac{\hbar I_0}{2e\phi_0}
\left(
1-\frac{5a\pi^4b^4}{\phi_0^4}
\right)
\end{equation}

and $K_4$ is 

\begin{equation}
K_4
=
\frac{5a\pi^2b^2\,\hbar I_0}{2e\,\phi_0^5}
\end{equation}
etc.

\section{\label{apdx:GL}Current--Phase Relationship of a Superconducting Nanowire from Ginzburg--Landau Theory}

In this appendix, we derive the current--phase relationship (CPR) of a uniform superconducting nanowire using one-dimensional Ginzburg--Landau (GL) theory. The derivation is valid near the superconducting critical temperature \(T_c\), where GL theory applies, and assumes a homogeneous wire carrying a uniform supercurrent, neglecting phase-slip processes.

\subsection{Ginzburg--Landau Free Energy}

We consider a superconducting wire oriented along the \(x\)-direction. The GL free-energy density is
\begin{equation}
f
=
\alpha |\psi|^2
+ \frac{\beta}{2}|\psi|^4
+ \frac{\hbar^2}{2m^\ast}
\left|
\left(\partial_x - i\frac{2e}{\hbar}A_x\right)\psi
\right|^2
\label{eq:GL_free_energy}
\end{equation}
where \(\psi(x)\) is the superconducting order parameter, \(m^\ast\) is the effective mass for superconducting electrons, and \(A_x\) is the vector potential.

We write the order parameter in amplitude--phase form,
\begin{equation}
\psi(x)=|\psi|\,e^{i\theta(x)}
\label{eq:order_parameter}
\end{equation}
and define the gauge-invariant phase gradient
\begin{equation}
q \equiv \partial_x\theta - \frac{2e}{\hbar}A_x
\label{eq:q_def}
\end{equation}

Assuming a uniform wire with constant current, both \(|\psi|\) and \(q\) are spatially uniform.

\subsection{Condensate Suppression by Superflow}

With the assumptions above, the gradient term in Eq.~\ref{eq:GL_free_energy} reduces to
\begin{equation}
\frac{\hbar^2}{2m^\ast}|\psi|^2 q^2
\label{eq:gradient_term}
\end{equation}

Minimizing the free-energy density with respect to \(|\psi|\) yields
\begin{equation}
\alpha + \beta|\psi|^2 + \frac{\hbar^2}{2m^\ast}q^2 = 0
\label{eq:GL_minimization}
\end{equation}

Defining the equilibrium condensate density
\begin{equation}
|\psi_0|^2 \equiv -\frac{\alpha}{\beta}
\label{eq:psi0}
\end{equation}
and the GL coherence length
\begin{equation}
\xi^2 \equiv \frac{\hbar^2}{2m^\ast|\alpha|}
\label{eq:xi_def}
\end{equation}
Eq.~\ref{eq:GL_minimization} gives
\begin{equation}
|\psi|^2 = |\psi_0|^2\left(1-\xi^2 q^2\right)
\label{eq:psi_q}
\end{equation}

This expression shows explicitly that the superconducting condensate is depleted by superflow.

\subsection{Supercurrent Density}

The supercurrent density in one dimension is
\begin{equation}
j_s = \frac{2e\hbar}{m^\ast}|\psi|^2 q
\label{eq:supercurrent_density}
\end{equation}

Substituting Eq.~\ref{eq:psi_q} into Eq.~\ref{eq:supercurrent_density}, we obtain
\begin{equation}
j_s(q)
=
\frac{2e\hbar}{m^\ast}|\psi_0|^2
\, q \left(1-\xi^2 q^2\right)
\label{eq:js_q}
\end{equation}

This is the GL depairing-current relation: the current increases linearly with \(q\) at small phase gradients and is suppressed at large \(q\) due to suppression of the condensate density (suppression of superconductivity by the current).

\subsection{Current--Phase Relationship}

For a wire of length \(L\), the total gauge-invariant phase difference across the wire is
\begin{equation}
\phi = \int_0^L q\,dx = qL
\label{eq:phase_drop}
\end{equation}
Thus,
\begin{equation}
q = \frac{\phi}{L}
\label{eq:q_phi}
\end{equation}

Let \(S\) be the cross-sectional area of the wire. The total supercurrent is \(I = S j_s\), which yields
\begin{equation}
I(\phi)
=
\frac{2e\hbar S}{m^\ast}|\psi_0|^2
\left(\frac{\phi}{L}\right)
\left(1-\xi^2\frac{\phi^2}{L^2}\right)
\label{eq:I_phi_raw}
\end{equation}

Introducing the characteristic phase scale
\begin{equation}
\phi_0 \equiv \frac{L}{\xi}
\label{eq:phi_c}
\end{equation}
The CPR can be written in compact form as
\begin{equation}
I(\phi)
=
I_0
\left(\frac{\phi}{\phi_0}\right)
\left[1-\left(\frac{\phi}{\phi_0}\right)^2\right]
\label{eq:CPR_GL}
\end{equation}
where
\begin{equation}
I_0 \equiv \frac{2e\hbar S}{m^\ast\xi}|\psi_0|^2
\label{eq:I0_def}
\end{equation}

Equation~\ref{eq:CPR_GL} is the GL current--phase relationship of a superconducting nanowire. It is odd in \(\phi\), approximately linear at small phase differences, and exhibits a maximum corresponding to the depairing critical current. Note that the relative contribution of the quartic term is inversely proportional to $\phi_0^2$, i.e., it is inversely proportional to the squared length of the wire $L$. Thus, longer wires will generate a low value of the qubit anharmonicity. In the main part of the paper, we discuss how a larger anharmonicity can be achieved by making a SQUID with two nanowires.

\subsection{Critical Current}

The maximum of Eq.~\ref{eq:js_q} occurs at
\begin{equation}
q\xi = \frac{1}{\sqrt{3}}
\label{eq:q_critical}
\end{equation}
which corresponds to a phase difference
\begin{equation}
\phi_c= \frac{\phi_0}{\sqrt{3}}
\label{eq:phi_critical}
\end{equation}

The resulting depairing critical current is
\begin{equation}
I_c
=
\frac{4e\hbar S}{3\sqrt{3}\,m^\ast\xi}|\psi_0|^2 =\frac{2}{3\sqrt3}I_0
\label{eq:Ic}
\end{equation}

This CPR differs qualitatively from the sinusoidal Josephson-junction CPR and reflects the distributed nature of superflow and condensate depletion in superconducting nanowires.

\section{\label{apdx:zpf}Anharmonicity from Zero-Point Fluctuations}

In this appendix, we derive the absolute and relative anharmonicities using a zero-point fluctuation formulation. We will also compare the results with Appendix A.

\subsection{Model Hamiltonian}

We consider a one-dimensional quantum oscillator whose potential energy deviates weakly from harmonicity due to a quartic term. The Hamiltonian is given by
\begin{equation}
\hat H
= \frac{\hat p^{\,2}}{2m_1}
+ \frac{1}{2} m_1 \omega^2 \hat x^{\,2}
- K_4 \hat x^{\,4}
\label{eq:H_full}
\end{equation}
where $\hat x =x$ is the coordinate operator which equals the coordinate itself, $m_1$ is the mass of the oscillator, $\omega$ is the harmonic frequency, and $K_4>0$ a softening anharmonicity parameter. The quartic (fourth power) term is assumed to be sufficiently small such that perturbation theory applies to the low-lying eigenstates.

\subsection{Zero-Point Fluctuations}

The natural length scale of the problem is set by the ground-state fluctuations of the harmonic oscillator. Introducing ladder operators, the position operator can be written as
\begin{equation}
\hat x = x_{\mathrm{zpf}}\left(\hat a+\hat a^\dagger\right)
\label{eq:x_operator}
\end{equation}
where the zero-point fluctuation amplitude is
\begin{equation}
x_{\mathrm{zpf}} = \sqrt{\frac{\hbar}{2m_1\omega}}
\label{eq:x_zpf}
\end{equation}

Physically, $x_{\mathrm{zpf}}$ characterizes the spatial extent of the ground-state wavefunction and therefore determines how strongly the oscillator samples nonlinear terms in the potential.

Suppose we write the Hamiltonian in a more generic form as

\begin{equation}
\hat H
= K_1\hat p^{\,2}
+ K_2x^2
- K_4x^4
\label{eq:H_full1}
\end{equation}
where $K_1=1/2m_1$ and $K_2=m_1\omega^2/2$.
With this generic notation, the frequency of classical oscillations is
\begin{equation}
\omega^2=4K_1K_2
\end{equation} and zero point fluctuation is
\begin{equation}
x_{\mathrm{zpf}}^4 = \frac{\hbar^2}{4}\frac{K_1}{K_2}
\label{eq:x_zpf1}
\end{equation}

\subsection{Dimensionless Anharmonicity Parameter}

Substituting Eq.~\ref{eq:x_operator} into the quartic term of Eq.~\ref{eq:H_full}, the nonlinear contribution to the Hamiltonian becomes
\begin{equation}
- K_4 \hat x^{\,4}
= -K_4 x_{\mathrm{zpf}}^{\,4} \left(\hat a+\hat a^\dagger\right)^4=-\hbar\omega\lambda \left(\hat a+\hat a^\dagger\right)^4
\label{eq:quartic_ladder}
\end{equation}
where we have introduced the dimensionless anharmonicity parameter
\begin{equation}
\lambda \equiv \frac{K_4 x_{\mathrm{zpf}}^{\,4}}{\hbar\omega}
\label{eq:lambda_def}
\end{equation}

This expression makes explicit that anharmonic effects scale with the fourth power of the zero-point fluctuations. Consequently, even a modest quartic coefficient $K$ can lead to appreciable anharmonicity if $x_{\mathrm{zpf}}$ is large.

\subsection{Energy Spectrum to First Order in $\lambda$}

Treating the quartic term as a perturbation, the energy of the $j$th eigenstate is ($j$ is a nonnegative integer)
\begin{equation}
E_j
= \hbar\omega\left(j+\tfrac12\right)
- \hbar\omega\,\lambda\left(6j^2+6j+3\right)
\label{eq:En_general}
\end{equation}
valid to first order in $\lambda$.

The three lowest energy levels are therefore
\begin{align}
E_0 &= \tfrac12\hbar\omega - 3\hbar\omega\,\lambda=\tfrac12\hbar\omega - 3K_4x_{zpf}^4
\label{eq:E0} \\
E_1 &= \tfrac32\hbar\omega - 15\hbar\omega\,\lambda=\tfrac32\hbar\omega - 15K_4x_{zpf}^4
\label{eq:E1} \\
E_2 &= \tfrac52\hbar\omega - 39\hbar\omega\,\lambda=\tfrac52\hbar\omega - 39K_4x_{zpf}^4
\label{eq:E2}
\end{align}

The normalized energy levels are defined as $\epsilon=E/\hbar\omega$. These levels are 
\begin{align}
\epsilon_0 &= \tfrac12- 3\lambda
\label{eq:E0a} \\
\epsilon_1 &= \tfrac32 - 15\lambda
\label{eq:E1a} \\
\epsilon_2 &= \tfrac52- 39\lambda
\label{eq:E2}
\end{align}

\subsection{Anharmonicity}

The anharmonicity is defined as the deviation from equal level spacing,
\begin{equation}
\alpha \equiv (E_2 - E_1) - (E_1 - E_0)
= E_2 - 2E_1 + E_0
\label{eq:alpha_def1}
\end{equation}

The relative anharmonicity is defined as:
\begin{equation}
  \alpha_{\text{r}} = \frac{\alpha}{E_1 - E_0} = \frac{E_2 - 2E_1 + E_0}{E_1 - E_0} = \frac{\epsilon_2 - 2\epsilon_1 + \epsilon_0}{\epsilon_1 - \epsilon_0}
\end{equation}

Substituting Eqs.~\ref{eq:E0}--\ref{eq:E2} into Eq.~\ref{eq:alpha_def1}, the harmonic contributions cancel identically, yielding
\begin{equation}
\alpha = -12\,\hbar\omega\,\lambda
\label{eq:alpha_lambda}
\end{equation}

Using the definition of $\lambda$ in Eq.~\ref{eq:lambda_def}, the anharmonicity can be expressed directly in terms of physical parameters as
\begin{equation}
\alpha = -12\,K_4\,x_{\mathrm{zpf}}^{\,4}
\label{eq:alpha_K}
\end{equation}

If $\lambda$ is small then $E_1-E_0\approx\hbar\omega$. Then the relative anharmonicity is $\alpha_r=-12\lambda=-12\frac{K_4 x_{\mathrm{zpf}}^{\,4}}{\hbar\omega}=-6\frac{K_4 x_{\mathrm{zpf}}^{\,4}}{\hbar\sqrt{K_1K_2}}=-\frac{3\hbar\sqrt{K_1} K_4}{2K_2\sqrt{K_2}}$. Comparing this result to that of Appendix~\ref{apdx:perturbation}, letting $4E_c/\hbar^2 = K_1$, $\beta_1 = K_2$, and $\gamma = -K_4$, we arrive at the same result obtained in Appendix E, thereby validating the zero-point fluctuations formulation. 

\subsection{Anharmonicity from Power-Law Potential}

More generally, consider
\begin{equation}
\hat H
= \frac{\hat p^{\,2}}{2m_1}
+ \frac{1}{2}m_1\omega^2 \hat x^{\,2}
+ U_{\mathrm{anh}}(\hat x)
\qquad
U_{\mathrm{anh}}(x)=-K_n x^{n}
\label{eq:H_general_app}
\end{equation}
with $K_n$ small. Introducing the dimensionless coordinate operator \(\hat q\) via
\begin{equation}
\hat x = x_{\mathrm{zpf}}\,\hat q
\label{eq:x_q}
\end{equation}
The anharmonicity to first order becomes
\begin{equation}
\alpha
= -K_n\,x_{\mathrm{zpf}}^{\,n}
\Big(
\langle 2|\hat q^{\,n}|2\rangle
-2\langle 1|\hat q^{\,n}|1\rangle
+\langle 0|\hat q^{\,n}|0\rangle
\Big)
\label{eq:alpha_general}
\end{equation}

Because harmonic-oscillator probability densities are even functions of position, all diagonal moments of odd powers vanish. Consequently, for odd \(n\),
\begin{equation}
\alpha = 0 \qquad (\text{odd } n,\ \text{to first order in }K_n)
\label{eq:alpha_odd}
\end{equation}

For even integer powers \(n=2(m+1)\) with the integer \(m\ge 1\), the matrix-element combination in Eq.~\ref{eq:alpha_general} evaluates to a universal numerical factor, yielding
\begin{equation}
\alpha
=
-\,K_n\,x_{\mathrm{zpf}}^{\,2m+2}\,
\frac{(2m+2)!}{2^{\,m}(m-1)!}
\qquad m\ge 1
\label{eq:alpha_even}
\end{equation}

Equivalently, expressed directly in terms of \(n\),
\begin{equation}
\alpha
=
-\,K_n\,x_{\mathrm{zpf}}^{\,n}\,
\frac{n!}{2^{\,n/2-1}\left(\frac{n}{2}-2\right)!}
\qquad n\ \text{even},\ n\ge 4
\label{eq:alpha_n}
\end{equation}
    
For example, if the anharmonic term is quartic, then $n=4$ and $\alpha=-12K_4x_{zpf}^4$. This is an important result for the nanowire-SQUID qubit analysis since, in a nonzero magnetic field, we always find some nonzero quartic term in the potential energy. Higher-order terms might be relevant if the magnetic field is zero. Thus, if $n=6$ then $\alpha=-180K_6x_{zpf}^6=-180K_6(E_C/K_2)^{3/2}$, and if $n=8$ then $\alpha=-2520K_8x_{zpf}^8$, etc.

\section{\label{apdx:WKB}WKB Semi-Classical Approximation}

This appendix section will be a summary of the application of the WKB semi-classical approximation towards estimating the energy levels of the Hamiltonian  Eq.~\ref{eq:cubic_Hamiltonian}. The WKB (Wentzel-Kramers-Brillouin) semi-classical approximation is a technique for finding approximate solutions to quantum-mechanical problems when the potential energy varies slowly compared to the particle’s wavelength. In other words, the total energy should be larger than the potential energy. In this paper, we used this technique to estimate the eigenenergies of the ground, first excited, and second excited states. 

Using the Hamiltonian from Eq~\ref{eq:cubic_Hamiltonian}, the corresponding time-independent Schr\"odinger equation (SE) is:

\begin{equation}\label{eq:SWE}
    \left[-4E_c\partial^2_\phi + V(\phi) \right]\Psi(\phi) =  E\Psi(\phi)
\end{equation}

Here, the potential of the function is the following:

\begin{equation}
    V(\phi) = \frac{\hbar I_0}{2e}\left(\left[\frac{\phi_0^2 - 3\pi^2b^2}{\phi_0^3}\right]\phi^2  - \left[\frac{1}{2\phi_0^3}\right]\phi^4\right)
\end{equation}
 Next, the WKB approximation uses the wavefunction ansatz of the form 

\begin{equation}
    \Psi(\phi) = \exp(iS(\phi))
\end{equation}

Where $S(\phi)$ is proportional to the action. Letting $S(\phi) = S_0 + S_1 + S_2 + \dots$ and inserting the wavefunction ansatz into Eq.~\ref{eq:SWE}, we obtain that $S(\phi)$ (up to first order):

\begin{equation}
    S_0(\phi) = \pm\int_{\phi_1}^{\phi_2}\sqrt{\frac{E - V(\phi)}{4E_c}}
\end{equation}

To obtain bound states, we impose the Bohr-Sommerfeld quantization condition:

\begin{equation}\label{eq:apdx_WKB}
    \int_{\phi_1}^{\phi_2}\sqrt{\frac{E - V(\phi)}{4E_c}} = \left(j + \frac{1}{2}\right)\pi
\end{equation}

Where $V(\phi_1) = V(\phi_2) = E$ are defined as the points where the wavefunction hits the potential energy function. Using Eq.~\ref{eq:apdx_WKB}, our code numerically estimates the total energy $E$. Below is a table of numerically calculated eigenenergies at various magnetic fields, compared with the WKB-approximated eigenenergies. Here, we use $\phi_0 = 10\pi$, $I_0 = 10^{-6}$, and $E_c/h \approx 0.3$GHz. 

\begin{table}[h]
\centering
\begin{tabular}{|c|ccc|ccc|}
\hline
& \multicolumn{3}{c|}{Numerical} & \multicolumn{3}{c|}{WKB} \\
\hline
$b$ & $E_0$ & $E_1$ & $E_2$ & $E_0$ & $E_1$ & $E_2$ \\
\hline
0   & 0.00357 & 0.01070 & 0.01783 & 0.00357 & 0.01070 & 0.01784 \\
1   & 0.00351 & 0.01054 & 0.01756 & 0.00351 & 0.01054 & 0.01757 \\
2   & 0.00335 & 0.01004 & 0.01673 & 0.00335 & 0.01004 & 0.01673 \\
3   & 0.00305 & 0.00914 & 0.01524 & 0.00305 & 0.00914 & 0.01524 \\
4   & 0.00257 & 0.00772 & 0.01286 & 0.00257 & 0.00772 & 0.01286 \\
5   & 0.00178 & 0.00535 & 0.00891 & 0.00178 & 0.00535 & 0.00891 \\
5.5 & 0.00108 & 0.00325 & 0.00540 & 0.00108 & 0.00325 & 0.00541 \\
5.7 & 0.00056 & 0.00167 & 0.00277 & 0.00056 & 0.00167 & 0.00276 \\
\hline
\end{tabular}
\caption{Comparison of numerically computed and WKB eigenenergies for varying magnetic field $b$.}
\label{tab:combined_energies}
\end{table}

Based on the above table, the WKB calculated eigenenergies align with the numerically calculated eigenenergies. Therefore, we conclude that, based on this test, the numerically calculated eigenenergies are accurate.

\section{\label{apdx:perturbation}Perturbation Theory for Generalized Quartic Potential}

We briefly review the perturbation theory employed when approximating the absolute and relative anharmonicity for a general quartic potential. Such a Hamiltonian can be written as: 

\begin{equation}
    \hat{H} = -4E_c \frac{\partial^2}{\partial \phi^2} + \beta_1 \phi^2 + \gamma \phi^4
\end{equation}

Next, we define the basic ladder operators for this system: 

\begin{align}
    a &= \frac{1}{2}\left(\frac{\beta_1}{E_c}\right)^{1/4}\phi + \left(\frac{E_c}{\beta_1}\right)^{1/4}\partial_\phi \\
    a^\dagger &= \frac{1}{2}\left(\frac{\beta_1}{E_c}\right)^{1/4}\phi - \left(\frac{E_c}{\beta_1}\right)^{1/4}\partial_\phi
\end{align}

Where $[a,a^\dagger] = 1$. Then, the complete Hamiltonian can be written in terms of the ladder operators: 

\begin{equation}
    \hat{H} = 4\sqrt{\beta_1 E_c}\left(a^\dagger a + \frac{1}{2}\right) + \frac{E_c\gamma}{\beta_1}(a^\dagger +a)^{4}
\end{equation}

Using perturbation theory, the leading order correction to the energy states can be computed as:

\begin{equation}
    \frac{E_c\gamma}{\beta_1}\langle j |(a^\dagger +a)^{4} | j \rangle = \frac{E_c\gamma}{\beta_1}(6j^2 + 6j + 3)
\end{equation}

Therefore, the total energy can be computed as 

\begin{equation}
    E_m = 4\sqrt{\beta_1 E_c}\left(j + \frac{1}{2}\right) + \frac{E_c\gamma}{\beta_1}(6j^2 + 6j + 3)
\end{equation}

Here, we define the frequency $\omega$ of this Hamiltonian as 

\begin{equation}
    \omega = 4\sqrt{\beta_1 E_c}
\end{equation}

We can then compute the anharmonicities:

\begin{equation}
    \alpha = \frac{12\gamma E_c}{\beta_1}, \;\; \alpha_r \approx \frac{3\gamma\sqrt{E_c}}{\beta_1^{3/2}}
\end{equation}

Here, we assumed that $E_1 - E_0 \approx \hbar \omega$ for weak nonlinearity. 

\section*{References}
\bibliography{references}

@PREAMBLE{
 "\providecommand{\noopsort}[1]{}" 
 # "\providecommand{\singleletter}[1]{#1}%" 
}

@book{Bezryadin_book,
  author    = {{Alexey Bezryadin}},
  title     = {Superconductivity in Nanowires: Fabrication and Quantum Transport},
  publisher = {Wiley‐VCH Verlag GmbH \& Co. KGaA},
  year      = {2012},
  address   = {Leipzig, Weinheim, Singapore}
}

@article{Bezryadin-2000,
	author = {{Alexey Bezryadin} and {Chun Ning Lau} and {Michael Tinkham}},
	journal = {Nature},
	month = {6},
	number = {5729},
	pages = {971--974},
	title = {{Quantum suppression of
superconductivity in
ultrathin nanowires}},
	volume = {404},
	year = {2000},
}

@article{shor_algorithm,
author = {{Peter Shor}},
title = {Polynomial-Time Algorithms for Prime Factorization and Discrete Logarithms on a Quantum Computer},
journal = {SIAM Journal on Computing},
volume = {26},
number = {5},
pages = {1484-1509},
year = {1997},
}

@article{PhysRevB.70.214506,
  title = {Phase slips in superconducting films with constrictions},
  author = {Chu, Sang L. and Bollinger, A. T. and Bezryadin, A.},
  journal = {Phys. Rev. B},
  volume = {70},
  issue = {21},
  pages = {214506},
  numpages = {6},
  year = {2004},
  month = {Dec},
  publisher = {American Physical Society},
}

@article{grovers,
  title = {Quantum Mechanics Helps in Searching for a Needle in a Haystack},
  author = {Grover, Lov K.},
  journal = {Phys. Rev. Lett.},
  volume = {79},
  issue = {2},
  pages = {325--328},
  year = {1997},
  month = {Jul},
  publisher = {American Physical Society},
}

@article{bilayer,
  title = {High Kinetic Inductance in Platinum‐Coated Aluminum Nanobridge Interferometers},
  author = {Nazhestkin, I.A. and others},
  journal = {Advanced Engineering Materials},
  volume = {27},
  pages = {2402385},
  year = {2025},
}

@article{transmon,
  title = {Charge-insensitive qubit design derived from the Cooper pair box},
  author = {Koch, Jens and Yu, Terri M. and Gambetta, Jay and Houck, A. A. and Schuster, D. I. and Majer, J. and Blais, Alexandre and Devoret, M. H. and Girvin, S. M. and Schoelkopf, R. J.},
  journal = {Phys. Rev. A},
  volume = {76},
  issue = {4},
  pages = {042319},
  year = {2007},
  month = {Oct},
  publisher = {American Physical Society},
}

@article{PhysRevB.94.165128,
  title = {Decoherence and radiation-free relaxation in Meissner transmon qubit coupled to Abrikosov vortices},
  author = {{Jaseung Ku}  and {Zack Yoscovits} and {Alex Levchenko} and {James Eckstein} and {Alexey Bezryadin}},
  journal = {Phys. Rev. B},
  volume = {94},
  issue = {16},
  pages = {165128},
  numpages = {14},
  year = {2016},
  month = {Oct},
  publisher = {American Physical Society},
}

@article{RevModPhys.76.411,
  title = {The current-phase relation in Josephson junctions},
  author = {Golubov, A. A. and Kupriyanov, M. Yu. and Il'ichev, E.},
  journal = {Rev. Mod. Phys.},
  volume = {76},
  issue = {2},
  pages = {411--469},
  numpages = {0},
  year = {2004},
  month = {Apr},
  publisher = {American Physical Society},
}

@article{PhysRevB.82.134518,
  title = {Superconducting nanowires as nonlinear inductive elements for qubits},
  author = {Ku, Jaseung and Manucharyan, Vladimir and Bezryadin, Alexey},
  journal = {Phys. Rev. B},
  volume = {82},
  issue = {13},
  pages = {134518},
  numpages = {11},
  year = {2010},
  month = {Oct},
  publisher = {American Physical Society},
}

@article{vijay-2010,
	author = {Vijay, R. and Levenson-Falk, E. M. and Slichter, D. H. and Siddiqi, I.},
	journal = {Applied Physics Letters},
	month = {5},
	number = {22},
	title = {{Approaching ideal weak link behavior with three dimensional aluminum nanobridges}},
	volume = {96},
    pages={223112},
	year = {2010},
}

@article{PhysRevLett.106.110502,
  title = {Observation of Quantum Jumps in a Superconducting Artificial Atom},
  author = {Vijay, R. and Slichter, D. H. and Siddiqi, I.},
  journal = {Phys. Rev. Lett.},
  volume = {106},
  issue = {11},
  pages = {110502},
  numpages = {4},
  year = {2011},
  month = {Mar},
  publisher = {American Physical Society},
}

@article{faramarzi-2021,
	author = {Faramarzi, Farzad and Day, Peter and Glasby, Jacob and Sypkens, Sasha and Colangelo, Marco and Chamberlin, Ralph and Mirhosseini, Mohammad and Schmidt, Kevin and Berggren, Karl K. and Mauskopf, Philip},
	journal = {IEEE Transactions on Applied Superconductivity},
	month = {3},
	number = {5},
	pages = {1--5},
	title = {{Initial design of a W-Band Superconducting Kinetic Inductance QUBIT}},
	volume = {31},
	year = {2021},
}

@article{PhysRevLett.93.077003,
  title = {Decoherence in Josephson Phase Qubits from Junction Resonators},
  author = {Simmonds, R. W. and Lang, K. M. and Hite, D. A. and Nam, S. and Pappas, D. P. and Martinis, John M.},
  journal = {Phys. Rev. Lett.},
  volume = {93},
  issue = {7},
  pages = {077003},
  numpages = {4},
  year = {2004},
  month = {Aug},
  publisher = {American Physical Society},
}

@article{hopkins-2005,
	author = {Hopkins, David S. and Pekker, David and Goldbart, Paul M. and Bezryadin, Alexey},
	journal = {Science},
	month = {6},
	number = {5729},
	pages = {1762--1765},
	title = {{Quantum interference device made by DNA templating of superconducting nanowires}},
	volume = {308},
	year = {2005},
}

@article{PhysRevB.83.184503,
  title = {Cratered Lorentzian response of driven microwave superconducting nanowire-bridged resonators: Oscillatory and magnetic-field induced stochastic states},
  author = {Brenner, Matthew W. and Gopalakrishnan, Sarang and Ku, Jaseung and McArdle, Timothy J. and Eckstein, James N. and Shah, Nayana and Goldbart, Paul M. and Bezryadin, Alexey},
  journal = {Phys. Rev. B},
  volume = {83},
  issue = {18},
  pages = {184503},
  numpages = {18},
  year = {2011},
  month = {May},
  publisher = {American Physical Society},
}

@article{PhysRevLett.103.087003,
  title = {Optimizing Anharmonicity in Nanoscale Weak Link Josephson Junction Oscillators},
  author = {Vijay, R. and Sau, J. D. and Cohen, Marvin L. and Siddiqi, I.},
  journal = {Phys. Rev. Lett.},
  volume = {103},
  issue = {8},
  pages = {087003},
  numpages = {4},
  year = {2009},
  month = {Aug},
  publisher = {American Physical Society},
}

@article{RevModPhys.51.101,
  title = {Superconducting weak links},
  author = {Likharev, K. K.},
  journal = {Rev. Mod. Phys.},
  volume = {51},
  issue = {1},
  pages = {101--159},
  numpages = {0},
  year = {1979},
  month = {Jan},
  publisher = {American Physical Society},
}

@article{sun-2025,
	author = {Sun, Cliff and Bezryadin, Alexey},
	journal = {Nano Express},
	month = {8},
	number = {3},
	pages = {035014},
	title = {{Multiple-nanowire superconducting quantum interference devices: critical currents, symmetries, and vorticity stability regions}},
	volume = {6},
	year = {2025},
}

@article{nakamura-1999,
	author = {Nakamura, Y. and Pashkin, Yu. A. and Tsai, J. S.},
	journal = {Nature},
	month = {4},
	number = {6730},
	pages = {786--788},
	title = {{Coherent control of macroscopic quantum states in a single-Cooper-pair box}},
	volume = {398},
	year = {1999},
}

@article{collaborators-2024,
	author = {Google Quantum AI and collaborators},
	journal = {Nature},
	month = {12},
	number = {8052},
	pages = {920--926},
	volume = {638},
    title = {{Quantum error correction below the surface code threshold}},
	year = {2024},
}

@Article{10.21468/SciPostPhysLectNotes.31,
	title={{Bogoliubov quasiparticles in superconducting qubits}},
	author={{Leonid Glazman} and {Gianluigi Catelani}},
	journal={SciPost Phys. Lect. Notes},
    volume={31},
	pages={1-40},
	year={2021},
	publisher={SciPost},
}

@article{PhysRevApplied.21.024047,
  title = {Improved coherence in optically defined niobium trilayer-junction qubits},
  author = {{Alexander Anferov}  and {Kan-Heng Lee}  and {Fang Zhao} and {Jonathan Simon}  and {David Schuster}},
  journal = {Phys. Rev. Appl.},
  volume = {21},
  issue = {2},
  pages = {024047},
  numpages = {20},
  year = {2024},
  month = {Feb},
  publisher = {American Physical Society},
}

@article{wang-2026,
	author = {Wang, Danqing and Wu, Yufeng and Pieczulewski, Naomi and Garg, Prachi and Pace, Manuel C. C. and Bøttcher, Charlotte G. L. and Mazumder, Baishakhi and Muller, David A. and Tang, Hong X.},
	journal = {Nature Materials},
	month = {1},
	title = {{All-nitride superconducting qubits based on atomic layer deposition}},
	year = {2026},
}

@article{purmessur-2025,
	author = {Purmessur, Cheeranjeev and Chow, Kaicheung and Van Heck, Bernard and Kou, Angela},
	journal = {Nature Communications},
	month = {12},
	number = {1},
	pages = {11456},
	title = {{Operation of a high-frequency, phase-slip qubit}},
	volume = {16},
	year = {2025},
}

@article{
doi:10.1126/science.1175552,
author = {Vladimir E. Manucharyan  and Jens Koch  and Leonid I. Glazman  and Michel H. Devoret },
title = {Fluxonium: Single Cooper-Pair Circuit Free of Charge Offsets},
journal = {Science},
volume = {326},
number = {5949},
pages = {113-116},
year = {2009},
doi = {10.1126/science.1175552},
URL = {https://www.science.org/doi/abs/10.1126/science.1175552},
}

@article{yan-2016,
	author = {Yan, Fei and Gustavsson, Simon and Kamal, Archana and Birenbaum, Jeffrey and Sears, Adam P and Hover, David and Gudmundsen, Ted J. and Rosenberg, Danna and Samach, Gabriel and Weber, S and Yoder, Jonilyn L. and Orlando, Terry P. and Clarke, John and Kerman, Andrew J. and Oliver, William D.},
	journal = {Nature Communications},
	month = {11},
	number = {1},
	pages = {12964},
	title = {{The flux qubit revisited to enhance coherence and reproducibility}},
	volume = {7},
	year = {2016},
	doi = {10.1038/ncomms12964},
	url = {https://www.nature.com/articles/ncomms12964},
}

@article{vodolazov-2018,
	author = {Vodolazov, D Yu and Aladyshkin, A Yu and Pestov, E E and Vdovichev, S N and Ustavshikov, S S and Levichev, M Yu and Putilov, A V and Yunin, P A and El’kina, A I and Bukharov, N N and Klushin, A M},
	journal = {Superconductor Science and Technology},
	month = {8},
	number = {11},
	pages = {115004},
	title = {{Peculiar superconducting properties of a thin film superconductor–normal metal bilayer with large ratio of resistivities}},
	volume = {31},
	year = {2018},
	doi = {10.1088/1361-6668/aada2e},
	url = {https://doi.org/10.1088/1361-6668/aada2e},
}

@article{koblischka-2021,
	author = {Koblischka, Michael Rudolf and Koblischka-Veneva, Anjela},
	journal = {Nanomaterials},
	month = {7},
	number = {8},
	pages = {1970},
	title = {{Fabrication of superconducting nanowires using the template method}},
	volume = {11},
	year = {2021},
	doi = {10.3390/nano11081970},
	url = {https://doi.org/10.3390/nano11081970},
}

@misc{purkayastha2026tunableanharmonicitysninasnanowire,
      title={Tunable anharmonicity in Sn-InAs nanowire transmons beyond the short junction limit}, 
      author={Amrita Purkayastha and Amritesh Sharma and Param J. Patel and An-Hsi Chen and Connor P. Dempsey and Shreyas Asodekar and Subhayan Sinha and Maxime Tomasian and Mihir Pendharkar and Christopher J. Palmstrøm and Moïra Hocevar and Kun Zuo and Michael Hatridge and Sergey M. Frolov},
      year={2026},
      eprint={2603.26895},
      archivePrefix={arXiv},
      primaryClass={cond-mat.mes-hall},
      url={https://arxiv.org/abs/2603.26895}, 
}

@article{PRXQuantum.2.040313,
  title = {Calibration of Flux Crosstalk in Large-Scale Flux-Tunable Superconducting Quantum Circuits},
  author = {Dai, X. and Tennant, D.M. and Trappen, R. and Martinez, A.J. and Melanson, D. and Yurtalan, M.A. and Tang, Y. and Novikov, S. and Grover, J.A. and Disseler, S.M. and Basham, J.I. and Das, R. and Kim, D.K. and Melville, A.J. and Niedzielski, B.M. and Weber, S.J. and Yoder, J.L. and Lidar, D.A. and Lupascu, A.},
  journal = {PRX Quantum},
  volume = {2},
  issue = {4},
  pages = {040313},
  numpages = {22},
  year = {2021},
  month = {Oct},
  publisher = {American Physical Society},
  doi = {10.1103/PRXQuantum.2.040313},
  url = {https://link.aps.org/doi/10.1103/PRXQuantum.2.040313}
}

@article{neill-2018,
	author = {Neill, C. and Roushan, P. and Kechedzhi, K. and Boixo, S. and Isakov, S. V. and Smelyanskiy, V. and Megrant, A. and Chiaro, B. and Dunsworth, A. and Arya, K. and Barends, R. and Burkett, B. and Chen, Y. and Chen, Z. and Fowler, A. and Foxen, B. and Giustina, M. and Graff, R. and Jeffrey, E. and Huang, T. and Kelly, J. and Klimov, P. and Lucero, E. and Mutus, J. and Neeley, M. and Quintana, C. and Sank, D. and Vainsencher, A. and Wenner, J. and White, T. C. and Neven, H. and Martinis, J. M.},
	journal = {Science},
	month = {4},
	number = {6385},
	pages = {195--199},
	title = {{A blueprint for demonstrating quantum supremacy with superconducting qubits}},
	volume = {360},
	year = {2018},
	doi = {10.1126/science.aao4309},
	url = {https://doi.org/10.1126/science.aao4309},
}

@article{malevannaya-2025,
	author = {Malevannaya, Elizaveta I. and Polozov, Viktor I. and Ivanov, Anton I. and Matanin, Aleksei R. and Smirnov, Nikita S. and Echeistov, Vladimir V. and Moskalev, Dmitry O. and Mikhalin, Dmitry A. and Shirokov, Denis E. and Panfilov, Yuri V. and Ryzhikov, Ilya A. and Andriyash, Aleksander V. and Rodionov, Ilya A.},
	journal = {Applied Physics Reviews},
	month = {9},
	number = {3},
	title = {{An engineering guide to superconducting quantum circuit shielding}},
	volume = {12},
	year = {2025},
	doi = {10.1063/5.0250262},
	url = {https://doi.org/10.1063/5.0250262},
}

@article{mooij-2006,
	author = {Mooij, J. E. and Nazarov, Yu. V.},
	journal = {Nature Physics},
	month = {2},
	number = {3},
	pages = {169--172},
	title = {{Superconducting nanowires as quantum phase-slip junctions}},
	volume = {2},
	year = {2006},
}

@misc{bottcher2025transmonqubitrealizedexploiting,
      title={A transmon qubit realized by exploiting the superconductor-insulator transition}, 
      author={C. G. L. Bøttcher and E. Önder and T. Connolly and J. Zhao and C. Kvande and D. Q. Wang and P. D. Kurilovich and S. Vaitiekėnas and L. I. Glazman and H. X. Tang and M. H. Devoret},
      year={2025},
      eprint={2510.19983},
      archivePrefix={arXiv},
}
\end{document}